\documentclass[10pt,twoside,twocolumn,a4paper,fleqn,showpacs,showkeys,nofootinbib]{revtex4-1}

\usepackage{physics}
\usepackage[caption = false]{subfig}
\usepackage{amsmath}
\usepackage{graphicx}
\usepackage{hyperref}
\usepackage{color}
\usepackage{amssymb}
\usepackage{amsfonts}
\usepackage{lipsum,appendix}
\usepackage{amsthm}
\usepackage{natbib}
\usepackage{multirow}
\usepackage{hhline}
\begin{document} 
\baselineskip 10pt 

\title {Hot electron relaxation in normal state of iron pnictides: memory function approach}

\author{Luxmi Rani} \email{luxmiphyiitr@gmail.com}
	
\author{Cem Sevik} \email{csevik@eskisehir.edu.tr }

\affiliation{Mechanical Engineering Department, Eskisehir Technical University, Eskisehir, TR 26555, Turkey.}

\date{\today}



\newcommand{\be}{\begin{equation}} 
\newcommand{\ee}{\end{equation}} 
\newcommand{\bea}{\begin{eqnarray}} 
\newcommand{\eea}{\end{eqnarray}} 
\newcommand{\bqa}{\begin{eqnarray}}
\newcommand{\eqa}{\end{eqnarray}}
\newcommand{\bwt}{\begin{widetext}}
\newcommand{\ewt}{\end{widetext}}
\newcommand{\mb}{\mathbf}
\newcommand{\mc}{\mathcal}
\newcommand{\nn}{\nonumber \\}
\newcommand{\sig}{\sigma}
\newcommand{\dg}{\dagger}
\newcommand{\ua}{\uparrow}
\newcommand{\da}{\downarrow}
\newcommand{\al}{\alpha}
\newcommand{\bt}{\beta}
\newcommand{\del}{\delta}
\newcommand{\trho}{\tilde{\rho}}
\newcommand{\e}{\epsilon}
\newcommand{\g}{\gamma}
\newcommand{\Ga}{\Gamma}
\newcommand{\w}{\omega}
\newcommand{\ti}{\tilde} 
\newcommand{\tcr}{\textcolor{red}}
\newcommand{\tcg}{\textcolor{green}}
\newcommand{\tcb}{\textcolor{blue}}
\newcommand{\La}{\Large}
\newcommand{\bo}{\bold}
\newcommand{\ok}{\tcr{\checkmark}}
\newcommand{\wrong}{\xmark}    
\newcommand{\lt}{\left}    
\newcommand{\rt}{\right} 

\begin{abstract}
	This study leads to the investigation of the non-equilibrium electron relaxation in the normal state of iron pnictides. Here  we consider the relaxation of electrons due to their coupling with magnons and phonons in the metallic state of iron pnictides using the memory function approach. In the present model, electrons live at a higher temperature than that of the phonon and magnon baths, mimicking a non-equilibrium steady state situation. Further we analyze theoretically the generalized Drude scattering rate within the framework of Two Temperature Model and study the full frequency and temperature behavior for it. In zero frequency regime, the rate of electron-magnon scattering and electron-phonon scattering shows a linear temperature dependence at higher temperature values greater than Bloch-Gr\"{u}neisen temperature. Whereas at lower temperature values,  $T\ll\Theta_{BG}$, corresponding scattering rates follow the temperature behavior as ($1/\tau_{e-p} \varpropto T^3$) and ($1/\tau_{e-m} \varpropto T^{3/2}$), respectively. In the AC regime, we compute that  $1/\tau \propto \omega^2$ for $\omega\ll\omega_{BG}$ and for the values greater than the Bloch-Gr\"{u}neisen frequency, it is $\omega$-independent. Also, in lower frequency and zero temperature limit, we have observed the different frequency scale of electron-magnon and electron-phonon scattering i.e ($1/\tau \propto \omega^{3/2}$) and ($1/\tau\propto \omega^{3}$).  These results can be viewed with the pump-probe experimental setting for the normal state of iron pnictides.

\end{abstract}

\maketitle 

\section{Introduction}

 Non-equilibrium electron relaxation in metals has been studied for a long time since the arrival of femtosecond pump-probe spectroscopy\cite{Schoenlein,Falkovsky,Wong,Singh,NS,Das}. In this method, the photo excitation creates a non-equilibrium distribution of otherwise Fermi distributed electrons (excited electrons form a non-Fermi distribution for a very short time scale of the order of tens of femto seconds). These non-equilibrium electrons relax (among themselves) to produce a Fermi distribution, but at ELEVATED temperatures. This hot Fermi-Dirac distribution of electrons then further relax via electron-phonon or electron-magnon scattering, thereby transferring energy from hot electrons to phonon or magnon bath. This is so called the Two Temperature Model (TTM)\cite{Wong,Singh,NS,Das}. Here Magnons are bosons like phonon and follows the Bose-Eienstein statistics\cite{Mahan,Ziman}. 
 
 In the current study, we are interested in calculating the relaxation time scale of hot electrons in single band of Iron-based superconductors (IBSCs). We apply the powerful mathematical technique known as the memory function formalism to analyze the electronic transport in iron-pnictides. Since the last two decades, iron pnictides have triggered new vistas among the research community to explore the experimental and theoretical aspects of these systems due to their potential technological applications\cite{Hosono,Paglione,Rotter,SG}. Iron is known for its \textit{3d}-orbital electron originated ferromagnetic material, while iron pnictides are metallic with the co-existence of anti-ferromagnetic and superconducting phase, which further depends on doping level. IBSCs have a semi-metallic parent compound and show small electronic anisotropy. Iron-based superconductors possess a two-dimensional layered tetragonal crystal structure and their electronic structure shows a complex behavior due to the presence of multiband and multi-pockets over the Brillouin zone in the momentum space\cite{Luo}. We are familiar with the multi-orbital nature of these systems. In such systems, we can observe the intra and inter orbital excitations by irradiating them optically. Therefore, these hot electrons relax via electron-phonon and electron-magnon scattering. The present work forms a basis which can also be extended to study the hot-electron relaxation in three orbital model Hamiltonian of iron pnictides. 
 
 The imaginary part of the memory function defines the generalized Drude scattering rate\cite{GW,Kubo,Mori,KR}, inverse of it will give us a time scale on which hot electrons relax in IBSCs. The basic idea of the memory function approach is to study the time dependent correlation functions systematically in the case of many-body correlated systems. The merit of using Memory function is that it directly deals with the dynamical behavior of electronic transport\cite{Singh,GW,Zwanzig,Kubo,Mori}. There is good number of pump-probe experiments reported on iron-pnictides\cite{Patz,Avigo,Yang,Kumar,Wu,Stojchevska}. With our study, we could be able to compare our computed timescale with that obtained from optical experiments.
 
 In this connection, our findings are as follows:
 In the normal metal state of IBSCs with isotropic energy dispersion and equilibrium situation (means that both temperature baths are in equal state i.e. $T_e = T_p$ and $T_e = T_m$), the rate of electron-magnon scattering obeys the power-law temperature dependence $1/\tau_{e-m}$ proportional to $T^n$ with the exponent \textit{n} = 3/2. While the rate of electron-phonon scattering follows the cubic power-law temperature dependence ($1/\tau_{e-p} \varpropto T^3$) in the case of low phonon temperature over the Bloch-Gr\"{u}neisen temperature ($\Theta_{BG}$) and zero frequency regime. However, in both electron-magnon and electron-phonon study, at higher temperatures, corresponding scattering rate shows the linear temperature dependence ($1/\tau_{e-m}$ or $1/\tau_{e-p} \varpropto T$). The zero temperature behavior is also studied.

  This paper is organized as follows. In section \ref{sec:theory}, we discuss the Hamiltonian model with electron-magnon interaction \ref{sec:theoryEM} and electron-acoustic phonon interaction \ref{sec:theoryEP} for the normal state of two dimensional iron based superconductors and  using the memory function formalism to calculate the scattering rate in the presence of two temperature baths (electron and phonon), and (electron and magnon), respectively. Further, we explain the results analytically in different frequency and temperature regimes over the Bloch-Gr\"{u}neisen temperature $(\Theta_{BG})$. In section \ref{sec:numeric}, we shows the behavior of scattering rate due to hot electron relaxation via electron-phonon and electron-magnon scattering in IBSCs numerically in all the limiting cases. Finally, we summarize our results and present our conclusions.
  To simplify the calculations, it is customary to pass to the system of units in which the frequently appearing Boltzmann constant and Planck's constant are set to unity i.e. $k_B = 1$ and  $\hbar = 1$ throughout the paper.
  
\section{Theoretical Framework}
\label{sec:theory}
 \subsection{Electron-Magnon interaction study} 
 \label{sec:theoryEM}
To study the electron relaxation in normal state of iron pnictides, we consider the total Hamiltonian having three parts such as free electron ($H_{\text{e}}$), free magnon ($H_{\text{m}}$) and interacting part i.e electron-magnon ($H_{\text{em}}$):
\bea
H & = & H_{\text{e}} + H_{\text{m}} + H_{\text{em}}.
\label{eq:H} 
\eea
Here,
\bea
H_{\text{e}}&=& \sum_{\mb{k}\sigma} \epsilon_k c^\dagger_{\mb{k}\sigma}c_{\mb{k}\sigma},\\
H_{\text{m}}& = &\sum_{q}\omega_q\left(b^\dagger_q b_q +\frac{1}{2}\right),\\
H_{\text{em}}&=& \sum_{\mb{k},\, \mb{k'},\sigma}\left[ D_m(\mb{k}-\mb{k'})c^\dagger_{\mb{k}\sigma}c_{\mb{k'}\sigma}b_{\mb{k}-\mb{k'}} + \text{H.c.} \right].
\label{eq:H1} 
\eea

 Here, $c^\dagger_{\mb{k}\sigma} (c_{\mb{k}\sigma})$ and $b^\dagger_q(b_q)$ are electron and magnon creation (annihilation) operators, $\sigma$ is a spin, $\mb{k}$ and  $\mb{q}=\mb{k}-\mb{k'}$ are electron and magnon momentum, respectively. Consider, $\epsilon_k= k^2/2m $ is the isotropic quadratic energy dispersion term in two dimensional IBSCs. $m$ is the electronic mass. $D_m(\mb{k}-\mb{k'})$ is the electron-magnon matrix element. $\w_q = C_m q^2$ is the magnon energy where $C_m$ is the magnon velocity. H.c is the Hermitian conjugate term. In the next section, we consider a steady-state situation in which magnon temperature stays constant at $T_m$, and electron temperature stays constant at $T_e$. This situation can be experimentally created by a continuous laser excitation of the normal metallic state of iron pnictides.
 
\subsubsection {Expressions}
 To calculate the magnon scattering rate in isotropic two dimensional IBSCs case, we use the G\"otze-W\"olfle formalism \cite{Singh,Das,GW,Kubo}. In this formalism, Memory function is expressed as
\bea
M(z,T_m,T_e)& = &\frac{z \chi(z)}{\chi_0-\chi(z)} \simeq \frac{z\chi(z)}{\chi_0}\bigg(1 + \frac{\chi(z)}{\chi_0} +....\bigg) \nn
&&\simeq \frac{z\chi(z)}{\chi_0},
\label{eq:M1}
\eea
$\chi(z)$ is the Fourier transform of the current-current correlation function:
\bea 
\chi(z)= i\int_0^\infty e^{izt}\langle [J_1, J_1]\rangle dt.
\eea 
Here, $J_1=\Sigma_{k\sigma}(\vec{k}.\hat{n})c^\dagger_{\mb{k}\sigma}c_{\mb{k}\sigma}$ is the current density operator. $\hat{n}$ is the unit vector along the direction of current. Using the equation of motion (EOM) method, the expression is
\bea
M(z, T_m, T_e)=\frac{\langle\langle [J_1, H];[J_1, H]\rangle\rangle_{z=0}-\langle\langle [J_1, H];[J_1, H]\rangle\rangle_z}{z\chi_0}.\nn
\label{eq:M}
\eea
 Evaluating the commutator in equation \ref{eq:M} using Hamiltonian: 
\bea
[J_1, H]= \sum_{k,k'}[(\vec{k}-\vec{k'}).\hat{n}][D_m(\mb{k}-\mb{k'})c^\dagger_{\mb{k}\sigma} c_{\mb{k'}\sigma}b_{\mb{k}-\mb{k'}}-H.c.].
\label{eq:j}
\eea
After some lengthy but straightforward calculations, the memory function expression takes the following form:
\begin{widetext}
\bea
M(z, T_m, T_e)&=&\frac{1}{\chi_0}\sum_{kk'} \left|D_m(\mb{k}-\mb{k'})\right|^2[(\vec{k}-\vec{k'}).\hat{n}]^2\times[{f (1-f')(1+n))-f'(1-f)n}]\nn
&&\times\frac{1}{(\epsilon_{k}-\epsilon_{k'}-\omega_{q})} \left[\frac{1}{(\epsilon_{k}-\epsilon_{k'}-\omega_{q}+z)} + \frac{1}{(\epsilon_{k}-\epsilon_{k'}-\omega_{q}-z)}\right].
\label{eq:MT}
\eea
Here, $f=f(\epsilon_k, \beta_e)$ and $f'=f(\epsilon_{k'}, \beta_e)$ are the Fermi-Dirac distribution functions at different energies such as $\epsilon_{k}$ and $\epsilon_{k'}$ and, electron temperature $T_e={1}/{\beta_e}$. $n=n({\omega_{q}}, \beta_m)$ is the Bose-Einstein distribution function followed by the magnon living at a temperature $T_m ={1}/{\beta_m}$ and having magnon energy $\w_q$. $T_m$ is the magnon temperature. \textit{z} is the complex frequency. $z=\w+i\eta$  and $\eta\rightarrow0^+$. $M(z,T_m, T_e)= M^{\prime}(\w, T_m, T_e)+i M^{\prime\prime}(\w, T_m, T_e)$. We are interested in the scattering rate which is the imaginary part of the memory function (i.e.$ M^{\prime\prime}(\w, T_m, T_e)= 1/\tau(\w,T_m, T_e)$). The use of identity $lim_{\eta\rightarrow0}\frac{1}{a \mp i\eta} = \mathbb{P}(\frac{1}{a}) \pm i\pi\delta(a)$ transforms the expression (\ref{eq:MT}) into delta function form. Thus apply the limit we have
\bea
\frac{1} {\tau(\w, T_m, T_e)}& = &\frac{1}{2m^2}\frac{\pi}{\chi_0}\sum_{kk'} \left|D_{m}(\mb{k}-\mb{k'})\right|^2[(\vec{k}-\vec{k'}).\hat{n}]^2 \times[{f (1-f')(1+n))-f'(1-f)n}]\nn
&&\times\frac{1}{\w}\Bigg[\delta(\epsilon_k-\epsilon_{k'}- \omega_{q}+\w) -\delta(\epsilon_k-\epsilon_{k'}- \omega_{q}-\w)\Bigg].
\label{eq:ImM}
\eea
After simplification the above equation, we get the final expression:
\bea
\frac{1} {\tau(\w, T_m, T_e)}&=&\frac{1} {\tau_{e-m}^0}\int_0^{q_{BG}} dq \times \frac{q^2}{\sqrt{1-({q}/{2k_f})^2}} \nn
&&\times\Bigg\lbrace(1-\frac{\w_q}{\w})\left[ n(\beta_m, \w_q)-n(\beta_e, \w_q-\w)\right] +(\rm terms\, with\, \w\,\rightarrow -\w)...\Bigg\rbrace.
\label{eq:ImMf}
\eea
Here, $1/\tau_{e-m}^0 =  D_{m0}^2 / 32 \pi^3 \chi_0 k_F$ and $q_{BG}$ is the Bloch-Gr\"{u}neisen momentum i.e. the maximum momentum for the magnon excitations (i.e. $q_{BG} =2k_F$). $n({\beta_m,\omega_{q}})= 1/(e^{\beta_m\omega_{q}} -1)$ and $n({\beta_e,\omega_{q}-\omega})= 1/(e^{\beta_e(\omega_{q}-\omega)} -1)$. Where, $\chi_0$ represents the static limit of correlation function (i.e. $ \chi_0 = Ne/m $) and $k_F$ is the Fermi momentum. This is the general result of electron-magnon scattering rate which is valid for all temperatures and frequencies regimes. The normal state of iron-based pnictides is semi-mettalic and shows small Fermi surface ($k_F$) as compared to the Debye surface ($k_D$). Therefore, the critical temperature separating the low-$T$ and high -$T$ behavior of the resistivity known as Bloch-Gr\"{u}neisen temperature ($\Theta_{BG}$) is followed. To understand the full behavior in different temperatures and frequencies regime, we need to perform the integral numerically. However in various limiting cases, analytical results can be obtained and are discussed as follows.

\subsubsection{Limiting cases}
\textbf{Case-I}: Zero frequency case (DC limit i.e $\omega \rightarrow 0$)\\
Within this limit, curly bracket in equation (\ref{eq:ImMf}) reduces to \\
\bea
2 \lim_{\omega \to 0} \left[ n(\beta_m, \w_q) - \sum_{m=0}^{\infty} \omega ^{2l}\left\lbrace\frac{\partial^{2l}}{\partial\w_q^{2l}}n(\beta_e, \w_q) + \w_q \frac{\partial^{2l+1}}{\partial\w_q^{2l+1}}n(\beta_e, \w_q)\right\rbrace \right],
\eea

 Here we consider only $l=0$ i.e. 
\bea
\frac{1} {\tau(\w,T_m, T_e)}= \frac{1} {\tau_{e-m}^0}\int_0^{q_{BG}} dq \times \frac{q^2}{\sqrt{1-({q}/{2k_f})^2}} \bigg(n(\beta_m, \w_q) - n(\beta_e, \w_q) -\w_q n'(\beta_e, \w_q)\bigg).
\label{eq:ImMfdc}
\eea
Using relations $\w_q = C_m q^2$ and $\w_{BG}\simeq\Theta_{BG} = C_m (2k_F)^2$, defining $\frac{\w_q}{T_m}=x$, $\frac{\w_q}{T_e}=y$, the equation (\ref{eq:ImMfdc}) becomes,
\bea 
\frac{1} {\tau(\w,T_m, T_e)}& = &\frac{1}{\tau_{e-m}^0} \times q_{BG}^2 \bigg[\left(\frac{T_m}{\Theta_{BG}}\right)^{3/2}\int_0^{\frac{\Theta_{BG}}{T_m}} dx \times \frac{1}{e^x -1} \frac{\sqrt{x}}{\sqrt{1-\left(\frac{x T_m}{\Theta_{BG}}\right)}}+ \nn
&& \left(\frac{T_e}{\Theta_{BG}}\right)^{3/2}\int_0^{\frac{\Theta_{BG}}{T_e}} dy \times \frac{\sqrt{y}}{\sqrt{1-\left(\frac{y T_e}{\Theta_{BG}}\right)}}\times \bigg(\frac{y-1}{e^y -1} + \frac{y}{(e^y -1)^2}\bigg)\bigg]
\label{eq:ImMfdc1}
\eea
The detail study of general DC case is as follows:

Subcase (a): $ T_m, T_e \ll \Theta_{BG}$, i.e., when both the magnon temperature and electron temperature are lower than the Bloch-Gr\"{u}neisen temperature. Equation (\ref{eq:ImMfdc1}) gives
\bea
\frac{1} {\tau(T_m, T_e)}&=& \frac{1} {\tau_{e-m}^0}\times q_{BG}^2 \bigg[\frac{\sqrt{\pi}}{2}\zeta\left(\frac{3}{2}\right)\times \left(\frac{T_m}{\Theta_{BG}}\right)^{3/2}+ \left(\frac{3\sqrt{\pi}}{4}\zeta\left(\frac{5}{2}\right)-\frac{\sqrt{\pi}}{2}\zeta\left(\frac{3}{2}\right)\right)\times \left(\frac{T_e}{\Theta_{BG}}\right)^{3/2}\bigg]\nn
&=&\frac{1} {\tau_{e-m}^0} \times q_{BG}^2\bigg[A_1T_m^{3/2} + B_1 T_e^{3/2}\bigg].
\eea
Here $A_1= \frac{\sqrt{\pi}}{2}\zeta\left(\frac{3}{2}\right)\times\Theta_{BG}^{-3/2}$ and $B_1= \left(\frac{3\sqrt{\pi}}{4}\zeta\left(\frac{5}{2}\right)-\frac{\sqrt{\pi}}{2}\zeta\left(\frac{3}{2}\right)\right) \times\Theta_{BG}^{-3/2}$.
The Riemann zeta function is $\zeta(n) = \sum_{u=1}^{\infty}= 1/u^n$ for
integral numbers \textit{n}.\\

Subcase (b) In high temperature regime, $ T_m, T_e \gg \Theta_{BG}$, equation (\ref{eq:ImMfdc1}) reduces to
\bea
\frac{1} {\tau(T_m, T_e)}&=&\frac{1} {\tau_{e-m}^0} \times q_{BG}^2 \bigg[ \frac{8}{3} \frac{T_m}{\Theta_{BG}} + \frac{16}{15} \bigg]\nn
&=& \frac{1} {\tau_{e-m}^0}\times q_{BG}^2\bigg[A_2 T_m + B_2\bigg]
\eea
Here $A_2= \frac{8}{3}\Theta_{BG}^{-1} $ and $B_2= \frac{16}{15}$. It is notable here that the scattering rate is independent of electron temperature, and it only depends on the magnon temperature.\\

Subcase (c) $T_m\gg\Theta_{BG}, T_e\ll\Theta_{BG}$. In this regime scattering rate can be written as
\bea
\frac{1} {\tau(T_m, T_e)}&=&\frac{1} {\tau_{e-m}^0}\times q_{BG}^2 \bigg[ \frac{8}{3} \frac{T_m}{\Theta_{BG}} + \left(\frac{3\sqrt{\pi}}{4}\zeta\left(\frac{5}{2}\right)-\frac{\sqrt{\pi}}{2}\zeta\left(\frac{3}{2}\right)\right)\times \left(\frac{T_e}{\Theta_{BG}}\right)^{3/2}\bigg]\nn
&=& \frac{1} {\tau_{e-m}^0} \times q_{BG}^2\bigg[A_3 T_m + B_3 T_e^{3/2}\bigg]
\eea

Here $A_3=\frac{8}{3}\Theta_{BG}^{-1}$ and $B_3=B_1$. It is important to note that $1/\tau$ leads to the linear magnon temperature dependence in high temperature regime and shows the $T_e^{3/2}$- dependence below the magnon Bloch-Gr\"{u}neisen temperature.\\

Subcase (d) $T_m\ll\Theta_{BG}, T_e\gg\Theta_{BG}$. In this regime equation (\ref{eq:ImMfdc1}) gives
\bea
\frac{1} {\tau(T_m, T_e)}&=&\frac{1} {\tau_{e-m}^0} q_{BG}^2\bigg[ \frac{\sqrt{\pi}}{2}\zeta\left(\frac{3}{2}\right)\times \left(\frac{T_m}{\Theta_{BG}}\right)^{3/2} + \frac{16}{15} \bigg]\nn
&=& \frac{1} {\tau_{e-m}^0}  q_{BG}^2 \bigg[A_4 T_m^{3/2}+ B_4 \bigg]
\eea

Here $A_4= \frac{\sqrt{\pi}}{2}\zeta\left(\frac{3}{2}\right) \Theta_{BG}^{-3/2} $ and $B_4= \frac{16}{15}$. Hence
${1} /{\tau(T_m, T_e)}$ has $T_m^{3/2}$- dependence. Scattering rate is independent of the electron temperature. These results are tabulated in Table \ref{T:resultsEM}. For $T_m=T_e$ in the equilibrium situation, the scattering rate due to electron-magnon coupling leads to the famous Bloch- $T^{3/2}$ law.

\textbf{Case-II}: Finite frequency regimes\\
Here, we proceed with equation (\ref{eq:ImMf}) to study the frequency dependent behaviour of scattering rate in different regimes.

\textbf{Subcase (1)}: Consider higher frequency limit i.e. $\w >> \w_{BG}$, then equation (\ref{eq:ImMf}) becomes
\bea
\frac{1} {\tau(\w, T_m, T_e)}=\frac{1} {\tau_{e-m}^0}\int_0^{q_{BG}} dq \times \frac{q^2}{\sqrt{1-({q}/{2k_f})^2}} \Bigg\lbrace 2n(\beta_m, \w_q)-n(\beta_e, -\w)-n(\beta_e, \w) \Bigg\rbrace.
\eea
This can be simplified by setting $\frac{\w_q}{T_m}=x$, $\frac{\w}{T_e}=\xi$, then expression takes the form:
\bea
\frac{1} {\tau(\w, T, T_e)}&=&\frac{1} {\tau_{e-m}^0}\times q_{BG}^2 \left(\frac{T_m}{\Theta_{BG}}\right)^{3/2}\int_0^{\frac{\Theta_{BG}}{T_m}} dx \times \frac{\sqrt{x}}{\sqrt{1-\left(\frac{x T_m}{\Theta_{BG}}\right)}}\times \left(\frac{2}{e^x -1}-\frac{1}{e^{-\xi} -1}-\frac{1}{e^\xi -1}\right).
\label{eq:ImMfac}
\eea
This is the general case of scattering rate when the frequency is higher than the magnon Bloch-Gr\"{u}neisen frequency in both higher and lower temperature regimes. The results are summarized in Table \ref{T:resultsEM}. In the low-temperature limit, the first term is $T_m^{3/2}$-dependent while in the limit of high temperatures, scattering rate is linear $T_m$-dependent. Note that for both low and high T-limit, the second term is temperature independent. Therefore, scattering rate is dictated only by magnon temperature at higher frequency limit($\w >> \w_{BG}$).

\textbf{Subcase (2)}: Consider $\w =\w_{BG}$, in this limit equation (\ref{eq:ImMf}) reduces
\bea
\frac{1} {\tau(\w, T_m, T_e)}&=&\frac{2} {\tau_{e-m}^0}\int_0^{q_{BG}} dq \times \frac{q^2}{\sqrt{1-({q}/{2k_f})^2}} \times\Bigg\lbrace n(\beta_m, \w_q)-n(\beta_e, 2\w_q) \Bigg\rbrace.
\eea
For simplicity, using the relations $\frac{\w_q}{T_m}=x$, and $\frac{\w_q}{T_e}=y$, then we have
\bea
\frac{1} {\tau(\w,T_m, T_e)}& = &\frac{1} {\tau_{e-m}^0}\times q_{BG}^2\bigg[\left(\frac{T_m}{\Theta_{BG}}\right)^{3/2}\int_0^{\frac{\Theta_{BG}}{T_m}} dx \times \frac{1}{e^x -1} \frac{\sqrt{x}}{\sqrt{1-\left(\frac{x T_m}{\Theta_{BG}}\right)}} - \nn
&& \left(\frac{T_e}{\Theta_{BG}}\right)^{3/2}\int_0^{\frac{\Theta_{BG}}{T_e}} dy \times \frac{\sqrt{y}}{\sqrt{1-\left(\frac{y T_e}{\Theta_{BG}}\right)}}\frac{1}{e^{2y} -1}\bigg].
\label{eq:acEM}
\eea
To further simplify the above equation, below we study the low and high temperature regimes separately.
In the low temperature regime $T_m, T_e\ll\Theta_{BG}$, equation (\ref{eq:acEM}) becomes
\bea
\frac{1} {\tau(\w, T_m, T_e)}&=&\frac{1} {\tau_{e-m}^0} \times q_{BG}^2 \bigg[\frac{\sqrt{\pi}}{2}\zeta\left(\frac{3}{2}\right)\times \left(\frac{T_m}{\Theta_{BG}}\right)^{3/2}-\frac{\sqrt{2\pi}}{8}\zeta\left(\frac{3}{2}\right)\times \left(\frac{T_e}{\Theta_{BG}}\right)^{3/2}\bigg]\nn
&=& \frac{1} {\tau_{e-m}^0}\times q_{BG}^2 \bigg[A_7 T^{3/2} + B_7T_e^{3/2}\bigg]
\eea
Here, $A_7=\frac{\sqrt{\pi}}{2}\zeta\left(\frac{3}{2}\right)(\Theta_{BG})^{-3/2}$  and $B_7 =- \frac{\sqrt{2\pi}}{8}\zeta\left(\frac{3}{2}\right)(\Theta_{BG})^{-3/2}$. 
In the high temperature regime $T_m, T_e\gg \Theta_{BG}$, equation (\ref{eq:acEM}) takes the following form
\bea
\frac{1} {\tau(\w, T_m, T_e)}&=&\frac{1} {\tau_{e-m}^0} \times q_{BG}^2\bigg[\frac{4}{3}\frac{T_m}{\Theta_{BG}} -\frac{2}{3}\frac{T_e}{\Theta_{BG}} \bigg]\nn
&=&\frac{1} {\tau_{e-m}^0}\times q_{BG}^2 \bigg[A_8 T_m + B_8 T_e\bigg]
\eea
Here $A_6= \frac{4}{3}\Theta_{BG}^{-1}$ and $B_6 = - \frac{2}{3}\Theta_{BG}^{-1}$. It is also noticeable here that in both the cases ($\w \gg \w_{BG}$ and $\w= \w_{BG}$), $ \frac{1} {\tau(\w, T_m,T_e)}$ shows the frequency independent behavior.

\textbf{Subcase (3)}: At finite but lower frequency $\w \ll \w_{BG}$ case, with relation $\w_q=C_m q^2$ the equation (\ref{eq:ImMf}) becomes

\bea
\frac{1} {\tau(\w, T_m, T_e)}=\frac{1} {\tau_{e-m}^0}\int_0^{q_{BG}} dq \times \frac{q^2}{\sqrt{1-({q}/{2k_f})^2}}\bigg[\frac{1}{e^\frac{w_q}{T_m}-1}-\sum_{l=0}^{\infty} \omega ^{2l}\left(\frac{\partial^{2l}}{\partial\w_q^{2l}}\frac{1}{e^\frac{w_q}{T_e}-1} + \w_q \frac{\partial^{2l+1}}{\partial\w_q^{2l+1}}\frac{1}{e^\frac{w_q}{T_e}-1}\right)\bigg]
\label{eq:ImMfacl}
\eea
Further simplify by putting the variables $\frac{\w_q}{T_m}=x$, $\frac{\w_q}{T_e}=y$, and for $l=1$, we have
\bea
\frac{1} {\tau(\w, T_m, T_e)}&=&\frac{1} {\tau_{e-m}^0}\times q_{BG}^2\bigg[\left(\frac{T_m}{\Theta_{BG}}\right)^{3/2}\int_0^{\frac{\Theta_{BG}}{T_m}} dx \times \frac{1}{e^x -1} \frac{\sqrt{x}}{\sqrt{1-\left(\frac{x T_m}{\Theta_{BG}}\right)}} - \nn
&& \w^2T_e^{-1/2}\frac{1}{\Theta_{BG}^{3/2}}\int_0^{\frac{\Theta_{BG}}{T_e}} dy \times \frac{\sqrt{y}}{\sqrt{1-\left(\frac{y T_e}{\Theta_{BG}}\right)}}\nn &&
\times\left(n_y + 3n_y^2 +2n_y^3 - y \left(n_y +7n_y^2 +12n_y^3+6n_y^4\right)\right)\bigg].
\eea
Here $n_y ={1}/(e^y -1)$.
This is the general equation of the imaginary part of memory function when frequency is lower than the Bloch-Gr\"{u}neisen frequency. After simplification,  we can see the subsubcases results in the Table \ref{T:resultsEM}. Proceeding as in the zero temperature ($T_m, T_e\rightarrow0$) case at all frequencies, the curly bracket of the equation (\ref{eq:ImMf}) leads to $ \left(1-\frac{\w_q}{\w}\right)e^{\beta \w_q}$ at $\w>\w_q$, and $ \left(\frac{\w_q}{\w}-1 \right)e^{\beta \w}$ at $\w<\w_q$. Further simplify the integral over $q_{BG}$, we have
\bea
\frac{1} {\tau(\omega)}= \frac{1} {\tau_{e-m}^0}\frac{1}{C_m^{3/2}} \left\{
		\begin{array}{ll}
			\left(\frac{4}{15}\right) \w^{3/2},  & \w \ll \w_{BG} \\
			\left(\frac{\pi}{2}\omega_{BG}^{3/2}-\frac{3\pi}{8}\frac{\omega_{BG}^{5/2}}{\w} \right),  & \w \gg \w_{BG}. \\
		\end{array}
		\right.
		\label{eq:temp0m}
		\eea
	
It is notice here that at lower frequency ($\w \ll \w_{BG}$) in zero temperature limit, the electron-magnon scattering rate shows the $\omega^{3/2}$ contribution. At frequency higher than BG-frequency, $\frac{1} {\tau(\omega)}$ gets saturation behavior. In the next section, we continue the analytical calculations by taking account the electron-phonon scattering study in normal metallic state of iron based pnictides at non-equilibrium situation. 

\begin{center}
	\begin{table}
		\begin{tabular}{|l|l|l|}
			\hline
			No & Regimes & $ \dfrac{1}{\tau}$\\
			\hline
			\hline
			1 & $\w= 0;\,\,\,\,T_e, T_m\ll\Theta_{BG}$ & $A_1T_m^{3/2} + B_1 T_e^{3/2}.$\\
			\hline
			& $\w= 0;\,\,\,\, T_e, T_m\gg\Theta_{BG}$  & $A_2T_m+ B_2$. \\
			\hline
			& $\w= 0;\,\,\,\, T_m\gg\Theta_{BG}, T_e\ll\Theta_{BG}$  & $A_3T_m+B_3T_e^{3/2}$. \\
			\hline
			& $\w= 0;\,\,\,\, T_m\ll\Theta_{BG},T_e\gg\Theta_{BG}$  & $A_4T_m^{3/2}+ B_4$. \\
			\hline
			2  &  $\w \gg \w_{BG};\,\,\,\, T_e, T_m\gg\Theta_{BG}$ & $A_5T_m+ B_5$.\\
			\hline
			& $\w \gg \w_{BG};\,\,\,\, T_e, T_m\ll \Theta_{BG}$ & $A_6T_m^{3/2}+ B_6$. \\
			\hline
			3  &  $\w= \w_{BG};\,\,\,\, T_m, T_e\ll\Theta_{BG}$ & $A_7T_m^{3/2} + B_7 T_e^{3/2}$.\\
			\hline
			& $\w = \w_{BG};\,\,\,\, T_m, T_e \gg\Theta_{BG}$ & $A_8T_m + B_8 T_e$. \\
			\hline
			4 & $\w \ll \w_{BG};\,\,\,\, T_m, T_e \gg \Theta_{BG}$ & $A_9T_m + B_9 T_e + C_9 \w^2 T_e^{1/2}$.\\
			\hline
			& $\w \ll \w_{BG};\,\,\,\, T_m, T_e \ll \Theta_{BG}$ & $A_{10}T_m^{3/2} + B_{10} T_e^{3/2} + C_{10} \w^2 T_e^{-1/2}$. \\
			\hline
			& $\w \ll \w_{BG};\,\,\,\, T_m \ll\Theta_{BG}, T_e \gg \Theta_{BG}$ & $A_{11}T_m^{3/2} + B_{11} T_e + C_{11} \w^2 T_e^{1/2}$. \\
			\hline
			& $\w \ll \w_{BG};\,\,\,\, T_m \gg\Theta_{BG}, T_e \ll \Theta_{BG}$ & $A_{12}T_m+ B_{12} T_e^{3/2} + C_{12} \w^2 T_e^{-1/2}$. \\
			\hline
			
		\end{tabular}
		\caption{The results of electrical scattering rate due to the electron-magnon interactions in different limiting cases.} 
		\label{T:resultsEM}
	\end{table}
\end{center}

\end{widetext}

\subsection{Electron-Phonon interaction study} 
 \label{sec:theoryEP}
In this context, one can write the Hamiltonian for normal state of iron
pnictide within one band scenario in the
following form:
 \bea
 H & = & H_{\text{e}} + H_{\text{p}} + H_{\text{ep}}.
 \label{eq:H2} 
 \eea
 The different parts of Hamiltonian mentioned in the above equation are defined as
\bea
H_{\text{e}} &= & \sum_{\mb{k}\sigma} \epsilon_k c^\dagger_{\mb{k}\sigma}c_{\mb{k}\sigma},\\
H_{\text{p}}& = &\sum_{\tilde{q}} \omega_{\tilde{q}}\left(b^\dagger_{\tilde{q}} b_{\tilde{q}}+\frac{1}{2}\right),\\
H_{\text{ep}} & = & \sum_{\mb{k},\, \mb{k'},\sigma}\left[ D(\mb{k}-\mb{k'})c^\dagger_{\mb{k}\sigma}c_{\mb{k'}\sigma}b_{\mb{k}-\mb{k'}} + \text{H.c.} \right].
 \label{eq:H1'} 
\eea
Here, $c^\dagger_{\mb{k}\sigma} (c_{\mb{k}\sigma})$ and $b^\dagger_{\tilde{q}} (b_{\tilde{q}})$ are electron and phonon creation (annihilation) operators having  spin $\sigma$ and $\mb{k}$ and  $\mb{\tilde{q}}=\mb{k}-\mb{k'}$ are electron and phonon momentum respectively. $\epsilon_k= k^2/2m $ is the isotropic quadratic energy dispersion term for free electrons in IBSCs. H.c is the Hermitian conjugate term. $D(\mb{k}-\mb{k'})$ is the electron-phonon matrix element. The phonons are considered as acoustic having a dispersion of the form $\omega_{\tilde{q}} = c\tilde{q}$. $c$ is the sound velocity.

As discussed earlier, the scattering rate can be
computed using (\ref{eq:M}). Thus with the definitions
of the current density operator into (\ref{eq:M}) and the model Hamiltonian (\ref{eq:H2}), the imaginary part of the memory
function or scattering rate can be expressed
as \footnote{where, the current density operator commutes with the non-interacting parts of the Hamiltonian, the interacting part i.e electron-phonon gives
  \be
 [J_1, H]= \sum_{k,k'}[(\vec{k}-\vec{k'}).\hat{n}][D(\mb{k}-\mb{k'})c^\dagger_{\mb{k}\sigma} c_{\mb{k'}\sigma}b_{\mb{k}-\mb{k'}}-H.c.].
 \ee 
}, 
 
 \bea
\frac{1} {\tau(\w, T_p, T_e)}& = &\frac{1}{2m^2}\frac{\pi}{\chi_0}\sum_{kk'} \left|D(\mb{k}-\mb{k'})\right|^2[(\vec{k}-\vec{k'}).\hat{n}]^2\nn
&&\times[{f (1-f')(1+n))-f'(1-f)n}]\nn
&&\times\frac{1}{\w}\Bigg[\delta(\epsilon_k-\epsilon_{k'}- \omega_{\tilde{q}}+\w) \nn &&-\delta(\epsilon_k-\epsilon_{k'}- \omega_{\tilde{q}}-\w)\Bigg].
\label{eq:ImM'}
\eea
After simplification the above equation, we get the final expression:
\bea
\frac{1} {\tau(\w, T_p, T_e)}&=&\frac{1} {\tau_{e-p}^0}\int_0^{\tilde{q}_{BG}} d\tilde{q} \times \frac{{\tilde{q}}^2}{\sqrt{1-({\tilde{q}}/{2k_f})^2}} \nn
&&\times\Bigg\lbrace(1-\frac{\w_{\tilde{q}}}{\w})\left[ n(\beta_p, \w_{\tilde{q}})-n(\beta_e, \w_{\tilde{q}}-\w)\right] \nn && +(\rm terms\, with\, \w\,\rightarrow -\w)..\Bigg\rbrace.
\label{eq:ImMfep}
\eea
Where, $1/\tau_{e-p}^0 =  D_{0}^2 / 32 \pi^3 \chi_0 k_f $ and $\tilde{q}_{BG}$ is the Bloch-Gr\"{u}neisen momentum i.e. the maximum momentum for the phonon excitations.
Here, $f=f(\epsilon_k, \beta_e)$ and $f'=f(\epsilon_{k'}, \beta_e)$ are the FD- distribution functions at different energies such as $\epsilon_{k}$ and $\epsilon_{k'}$ and, electron temperature $T_e={1}/{\beta_e}$. $n=n(\w_{\tilde{q}}, \beta_p)$ is the Bose-Einstein distribution function obeyed by the phonons living at a temperature $T_p ={1}/{\beta_p}$ and having energy $\w_{\tilde{q}}$. $T_p$ is the phonon temperature. This is the final expression of frequency and temperature dependent electron-phonon scattering rate.\\

Further, the above expression in different frequency and temperature domains can be discussed or analyzed as follows:\\
 \begin{widetext}
\textbf{Case-I}: The zero frequency limit: $\omega \rightarrow 0$\\
Within this limit, curly bracket in equation (\ref{eq:ImMfep}) reduces to \\
\bea
2 \lim_{\omega \to 0} \left[ n(\beta_p, \w_{\tilde{q}}) - \sum_{m=0}^{\infty} \omega ^{2s}\left\lbrace\frac{\partial^{2s}}{\partial\w_{\tilde{q}}^{2s}}n(\beta_e, \w_{\tilde{q}}) + \w_q \frac{\partial^{2s+1}}{\partial\w_{\tilde{q}}^{2s+1}}n(\beta_e, \w_{\tilde{q}})\right\rbrace \right],
\eea

 Here we consider only $s=0$ i.e. 
\bea
\frac{1} {\tau(\w,T_p, T_e)}= \frac{2} {\tau_{e-p}^0}\int_0^{{\tilde{q}}_{BG}} d{\tilde{q}} \times \frac{{\tilde{q}}^2}{\sqrt{1-({{\tilde{q}}}/{2k_f})^2}} \bigg(n(\beta_p, \w_{\tilde{q}}) - n(\beta_e, \w_{\tilde{q}}) -\w_{\tilde{q}} n'(\beta_e, \w_{\tilde{q}})\bigg).
\label{eq:ImMfdcep}
\eea
Using the relation $\w_{\tilde{q}} =c \tilde{q}$ and defining $\frac{\w_{\tilde{q}}}{T_p}=x^\prime $, $\frac{\w_{\tilde{q}}}{T_e}=y^\prime$, the equation (\ref{eq:ImMfdcep}) becomes,
\bea 
\frac{1} {\tau(\w,T_p, T_e)}& = &\frac{2}{\tau_{e-p}^0} \times {\tilde{q}}_{BG}^3 \bigg[\left(\frac{T_p}{\Theta_{BG}}\right)^{3}\int_0^{\frac{\Theta_{BG}}{T_p}} dx^\prime \times \frac{x^{\prime 2}}{e^{x^\prime} -1} \frac{1}{\sqrt{1-\left(\frac{x^\prime T_p}{\Theta_{BG}}\right)^2}}+ \nn
&& \left(\frac{T_e}{\Theta_{BG}}\right)^{3}\int_0^{\frac{\Theta_{BG}}{T_e}} dy^\prime \times \frac{y^{\prime 2}}{\sqrt{1-\left(\frac{y' T_e}{\Theta_{BG}}\right)^2}}\times \bigg(\frac{y^\prime-1}{e^{y^\prime} -1} + \frac{y^\prime}{(e^{y^\prime} -1)^2}\bigg)\bigg]
\label{eq:ImMfdc_EP}
\eea
The above expression is the final result of electron-phonon scattering rate at zero frequency limit. Thus we get various limiting expressions which are summarized in Table \ref{T:resultsEP}.\\

\textbf{Case-II}: \textbf{Finite frequency regimes}\\
 In the finite frequency regimes, the asymptotic results of the
scattering rate in different temperature and the frequency regimes are shown in Table \ref{T:resultsEP}. It is important to note here that in zero temperature limit (i.e. $T_e \rightarrow 0, T_p \rightarrow 0$) and at lower frequency ($\w \ll \w_{BG}$), the electron-phonon scattering rate shows the $\omega^{3}$ contribution and at higher frequency ($\w \gg \w_{BG}$), $\frac{1}{\tau(\w)} \propto constant$. However, at lower frequency and finite temperature, $\frac{1} {\tau(\w,T_p, T_e)} $ follows $\w^2$-dependence with electron temperature coefficient. 
\end{widetext}
\bwt
\begin{center}
\begin{table}
\begin{tabular}{|l|l|l|}
\hline
	No & Regimes & $ \dfrac{1}{\tau}$\\
\hline
 \hline
	1 & $\w= 0;\,\,\,\,T_e, T_p\ll\Theta_{BG}$ & $a_1T_p^{3} + b_1 T_e^{3}.$\\
\hline
	 & $\w= 0;\,\,\,\, T_e, T_p\gg\Theta_{BG}$  & $a_2T_p+ b_2$. \\
\hline
	 & $\w= 0;\,\,\,\, T_p\gg\Theta_{BG}, T_e\ll\Theta_{BG}$  & $a_3T_p+b_3T_e^{3}$. \\
\hline
	 & $\w= 0;\,\,\,\, T_p\ll\Theta_{BG},T_e\gg\Theta_{BG}$  & $a_4T_p^{3}+ b_4$. \\
\hline
2  &  $\w \gg \w_{BG};\,\,\,\, T_e, T_p\gg\Theta_{BG}$ & $a_5T_p+ b_5$.\\
\hline
   & $\w \gg \w_{BG};\,\,\,\, T_e, T_p\ll \Theta_{BG}$ & $a_6T_p^{3}+ b_6$. \\
 \hline
3  &  $\w= \w_{BG};\,\,\,\, T_p, T_e\ll\Theta_{BG}$ & $a_7T_p^{3} + b_7 T_e^{3}$.\\
 \hline
 & $\w = \w_{BG};\,\,\,\, T_p, T_e \gg\Theta_{BG}$ & $a_8T_p + b_8 T_e$. \\
 \hline
4 & $\w \ll \w_{BG};\,\,\,\, T_p, T_e \gg \Theta_{BG}$ & $a_9T_p+ b_9 T_e + c_9 \w^2 T_e$.\\
\hline
& $\w \ll \w_{BG};\,\,\,\, T_p, T_e \ll \Theta_{BG}$ & $a_{10}T_p^{3} + b_{10} T_e^{3} + c_{10} \w^2 T_e$. \\
\hline
& $\w \ll \w_{BG};\,\,\,\, T_p \ll\Theta_{BG}, T_e \gg \Theta_{BG}$ & $a_{11}T_p^{3} + b_{11} T_e + c_{11} \w^2 T_e$. \\
\hline
& $\w \ll \w_{BG};\,\,\,\, T_p \gg\Theta_{BG}, T_e \ll \Theta_{BG}$ & $a_{12}T_p+ b_{12} T_e^{3} + c_{12} \w^2 T_e$. \\
\hline
\end{tabular}
\caption{The analytical results of scattering rate due to the electron-phonon interaction in different frequency and temperature regimes.}
 \label{T:resultsEP}
 \end{table}
\end{center}
\ewt


\section {Numerical analysis: }
\label{sec:numeric}

In order to analyze the non-equilibrium relaxation of hot electrons in the presence of magnons subsystem and phonons subsystem, we have numerically computed the general Eqn.\ref{eq:ImMf} and Eqn.\ref{eq:ImMfep} in different frequency and temperature regimes in single-band scenario for normal metal state of iron-pnictides.

In Figure \ref{fig:fig1}(a), we depict the magnon temperature dependence of scattering rate $1/\tau(T_m, T_e)$ normalized by $1/\tau_{e-m}^0$ at zero frequency and at different electron temperatures.  Plots of Fig.\ref{fig:fig1}(c) shows the phonon temperature dependence in scattering rate $1/\tau(T_p, T_e)$ normalized by $1/\tau_{e-p}^0$ in DC case. From Fig. \ref{fig:fig1}(a) and \ref{fig:fig1}(c), we observe that at high temperatures ($T_e, T\gg\Theta_{BG}$), $1/\tau(T, T_e)\propto T$. This can also be seen in the corresponding case ($T_e, T\gg\Theta_{BG}$) in Table \ref{T:resultsEM} and \ref{T:resultsEP}.
At very low temperature ($ T_m, T_e \ll \Theta_{BG}$), $1/\tau\propto T_m^{3/2}$  and $ T_e^{3/2}$ in electron-magnon scattering study. While, due to electron-phonon interaction, the scattering rate shows the cubic power law temperature dependence $1/\tau\propto T_p^{3}$  and $ T_e^{3}$.
Figures.\ref{fig:fig1}(b) and \ref{fig:fig1}(d) show the dependence of e-m scattering and e-ph scattering $1/\tau$ on $T_e$ in the DC limit. It is noticed that $1/\tau$  is independent of $T_e$ when $T_e \gg \Theta_{BG}$ in both the cases.
Contour plots (Fig.\ref{fig:fig1}(e), (f) and Fig.\ref{fig:fig1}(g), (h)) depict the constant value of $1/\tau$ in $T_p$- $T_e$ and $T_m$- $T_e$ plane. 

  In Figure\ref{fig:fig1}, from the contour plots, we observe that they are not symmetric around $T_m=T_e$ and $T_p=T_e$ line. The reason behind this asymmetry in DC case is that the scattering rate is differently effected by magnon temperature, phonon temperature, and electron temperature (the prefactor $A_1$ of $T_m^{3/2}$ term is not equal to the prefactor $B_1$ of $T_e^{3/2}$ terms in electron-magnon interaction study and the prefactor $a_1 \neq b_1$ of temperatures in electron-acoustic phonon coupling study). At very low temperature, $T^3$ or $T^{3/2}$ scaling behavior is due to Pauli blocking effect. We see  that at high temperature regime ($T_e, T_m\gg\Theta_{BG}$), $1/\tau(T_m, T_e)$ is proportional to magnon temperature $T_m$, not $T_e$. Similarly, we have noticed the linear phonon temperature dependence on electron-phonon scattering rate at higher temperature over $\Theta_{BG}$. The reason for this behavior is that at high temperatures Boson modes scale as $k_B T (<n_q>=\frac{1}{e^{\beta \w_q} -1} \propto k_B T)$, thus scattering increases with increasing temperatures linearly. For higher electron temperature case, the electron distribution can be explained as Boltzmann distributions because $\Theta_{BG} \simeq \theta_F$ (Fermi temperature). The temperature effect is exponentially reduced in this case as compared to magnons or phonons ($<n_q> \propto T$). Thus at high temperatures, the scattering rate shows linear temperature dependence T.

Figures \ref{fig:fig2}(a-d) depict the variations of the scattering rate at finite but lower frequency and different temperature regimes. From Figs. \ref{fig:fig2}(a) and \ref{fig:fig2}(c), it is observed that at lower magnon and lower phonon temperature range, the magnitude of scattering rate increases with increasing temperature as $T^n$ behavior (i.e. $n=3/2$ for electron-magnon and $n=3$ for electron-phonon scattering). At lower frequency and higher temperature limit, both scattering rates show T-linear behavior. 
In Figures \ref{fig:fig2}(b) and \ref{fig:fig2}(d), the variation of electron temperature dependence of $\tau_0/\tau(\w,T_m, T_e), \, and \, \tau_0/\tau(\w,T_p, T_e) $ at different magnon and  phonon temperatures scaled with corresponding Bloch-Gr\"{u}neisen temperature are shown. On increasing the magnon and phonon temperatures, corresponding scattering rate shows the saturation at higher electron temperature regime.  The insets of the figures \ref{fig:fig2}(a) and \ref{fig:fig2}(c) are more elaborated in the lower temperature regimes which show the deviations from linearity. These signatures are in accord with our
 analytical predictions Table \ref{T:resultsEM} and Table \ref{T:resultsEP}.
 
 We further analyzed the scattering rate at zero temperature limit (at all frequencies) in which electron relax when both magnon subsystem and phonon subsystem are at ultralow temperature. Compare scattering rate picture with frequency is shown in Figure \ref{fig:fig3}(a). It is observed that electron-magnon scattering rate is higher than electron-phonon scattering rate in lower frequency regime. At higher frequency domain, both e-m and e-p scattering rates get saturation. In Figure\ref{fig:fig3}(b), we depict the scattering rate due to bosons (magnons or phonons) interactions in low frequency domain. It is found that the scattering rate of magnons follows the relation of $\w^{3/2}$, while phonons scattering rate shows the more curvature in trend and seem to possess the analytic scaling $\w^3$.
 
 In Figure \ref{fig:fig4}, we plot the scattering rate $1/\tau_{e-m}$ and $1/\tau_{e-p}$, normalized by $1/\tau_{e-m}^0$ and $1/\tau_{e-p}^0$ with frequency at different temperatures. Figures \ref{fig:fig4}(a) and \ref{fig:fig4}(c) depict the variations of scattering rate with frequency at different electron temperatures and at fixed magnon temperature and phonon temperature, respectively, scaled with corresponding Bloch-Gr\"{u}neisen temperature. Scattering rate increases with increasing electron temperatures. Fig.\ref{fig:fig4}(b) and (d) show the e-m and e-p scattering variation with frequency, normalized by $\omega_{BG}$, at fixed electron temperature and different magnon and phonon temperatures, respectively. It is observed that the magnitude of scattering rate is changed due to the corresponding temperature dependent coefficients. At frequency lower than Bloch-Gr\"{u}neisen frequency, it follows $\w^2$-trend. These  temperature and frequency dependent results should be experimentally tested. \begin{widetext}

  \begin{figure}
	\centering
  	 \subfloat[]{\includegraphics[angle=0,width=5.64cm]{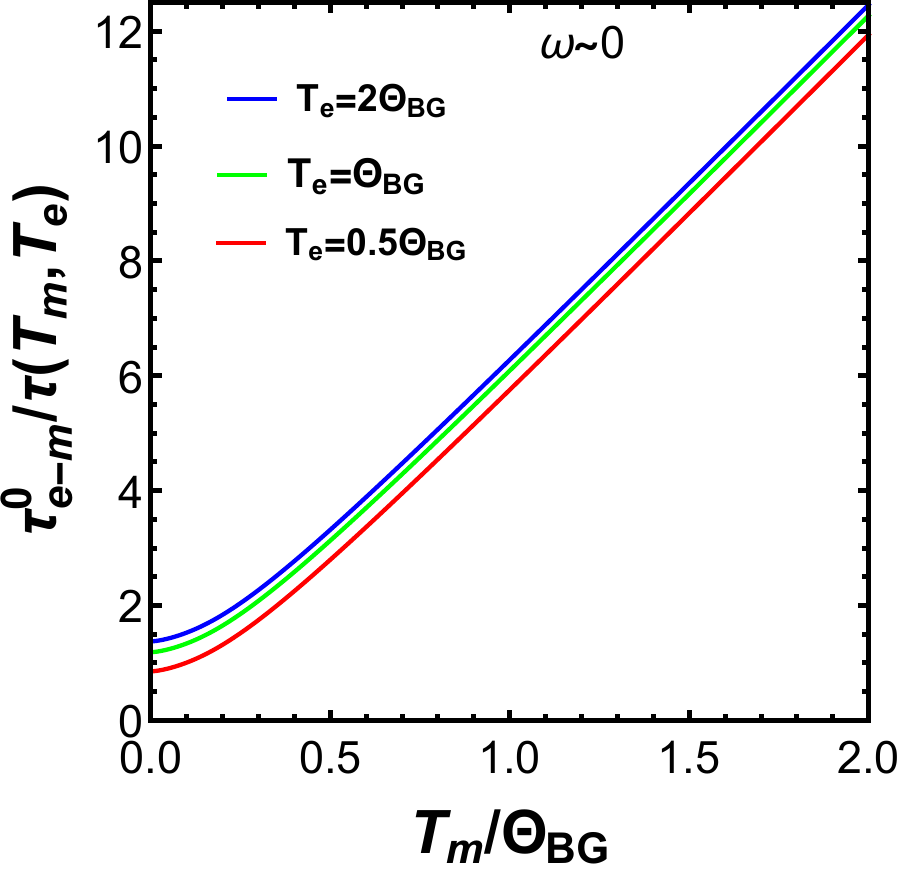}} 
	 \subfloat[]{\includegraphics[angle=0,width=5.64cm]{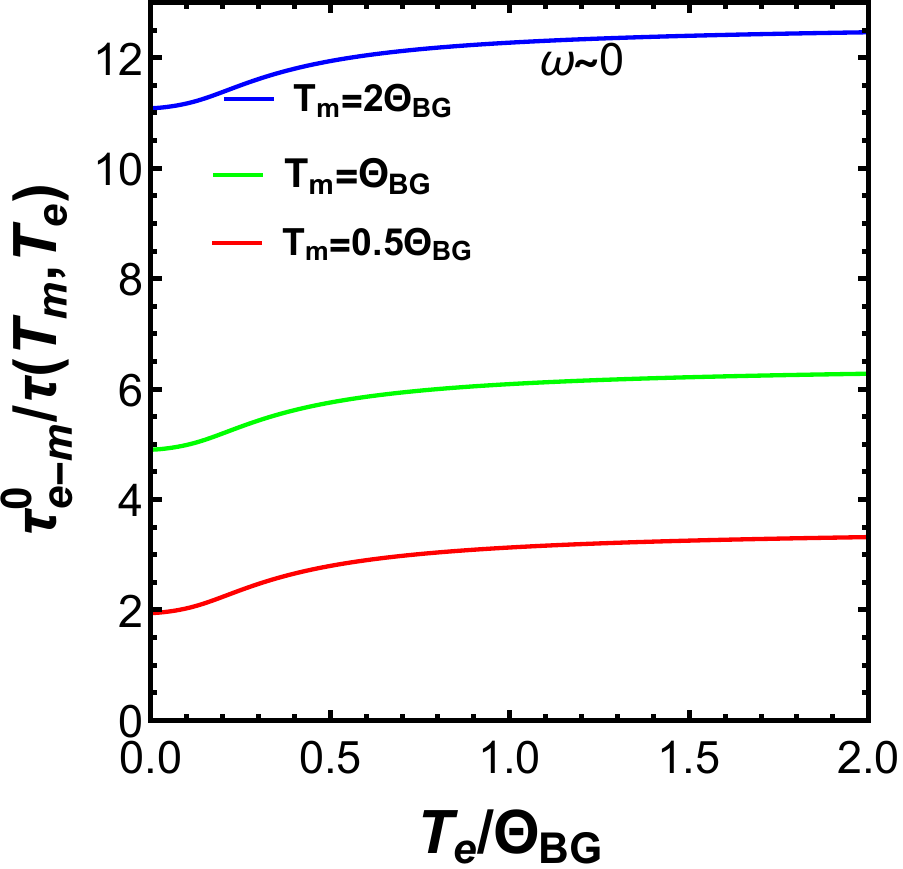}}
	 \subfloat[]{ \includegraphics[angle=0,width=5.64cm]{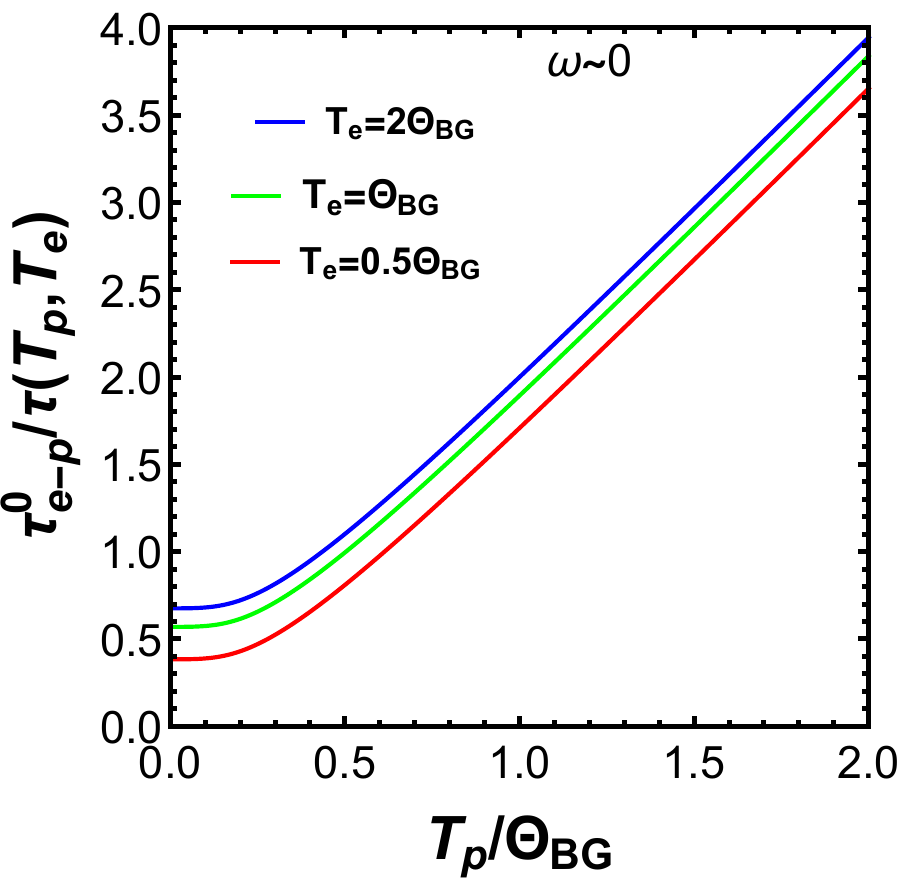}}\\
	 \subfloat[]{ \includegraphics[angle=0,width=5.64cm]{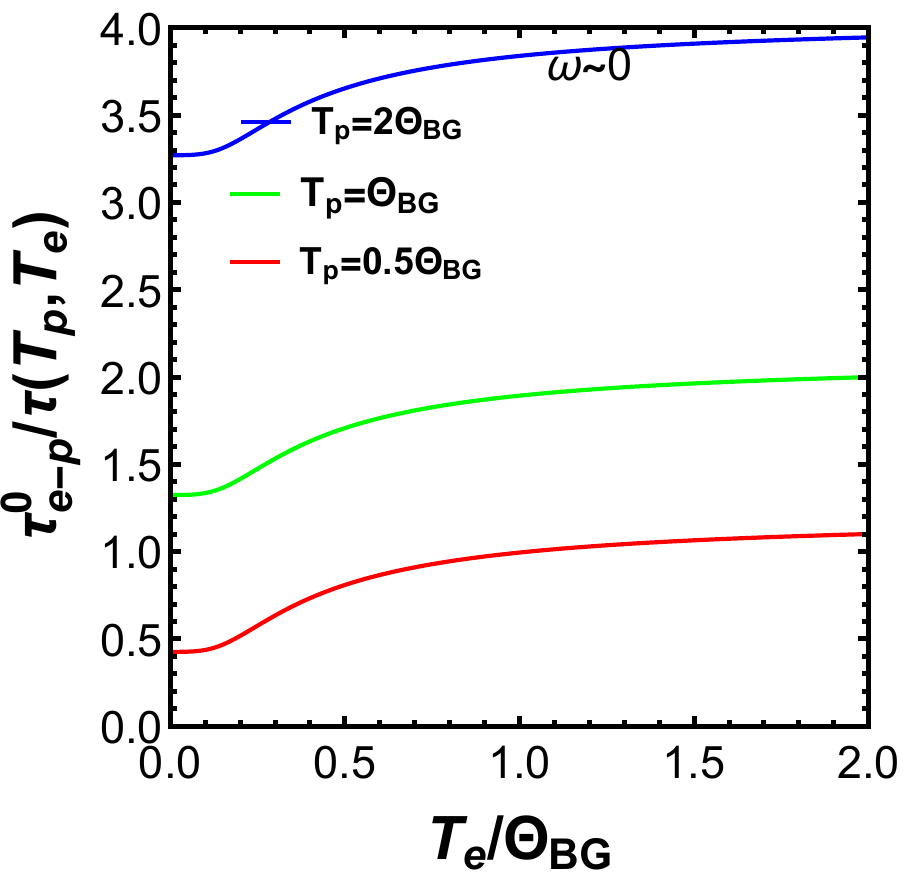}}
     \subfloat[]{\includegraphics[angle=0,width=5.8cm]{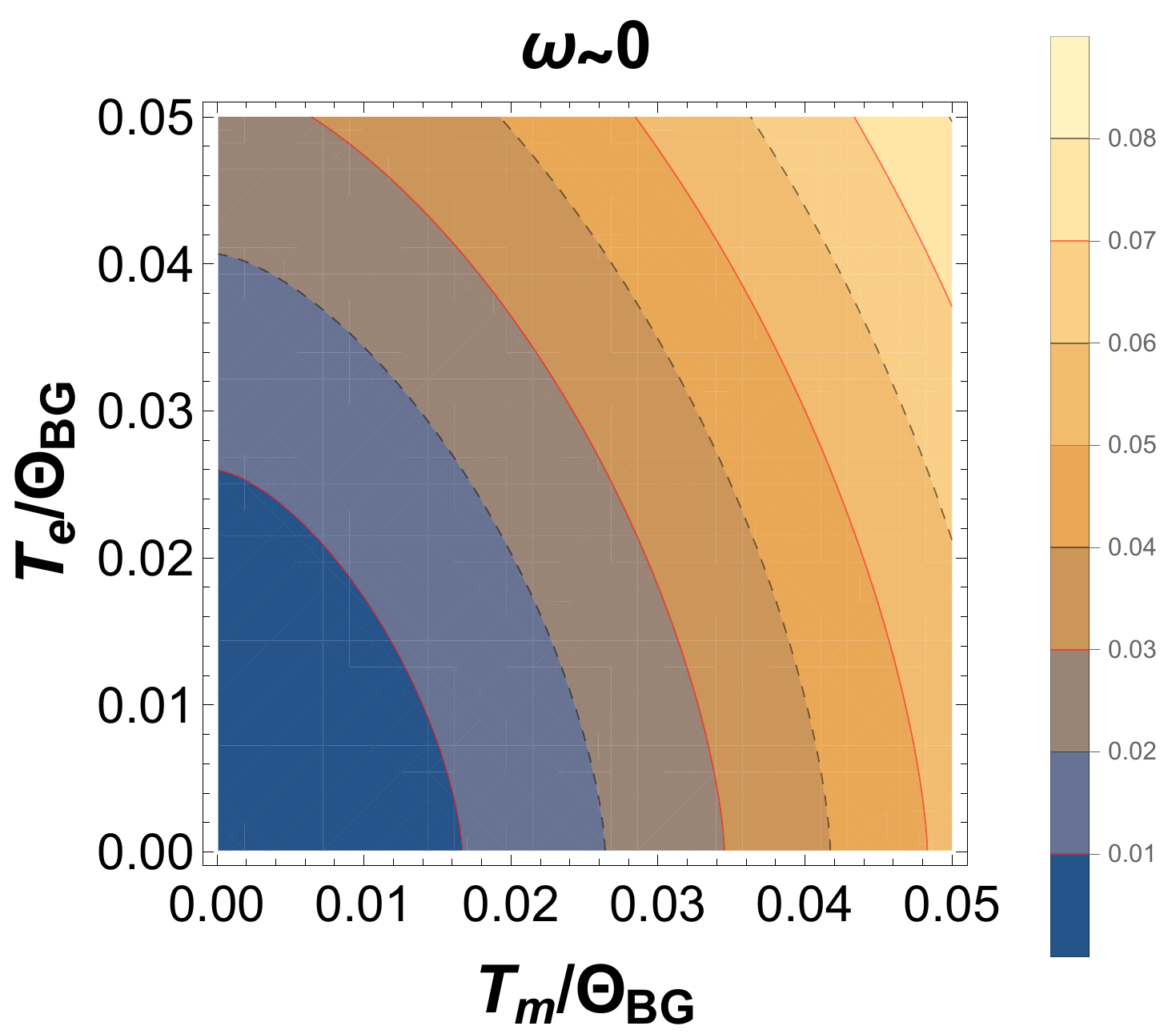}} 
     \subfloat[]{\includegraphics[angle=0,width=5.8cm]{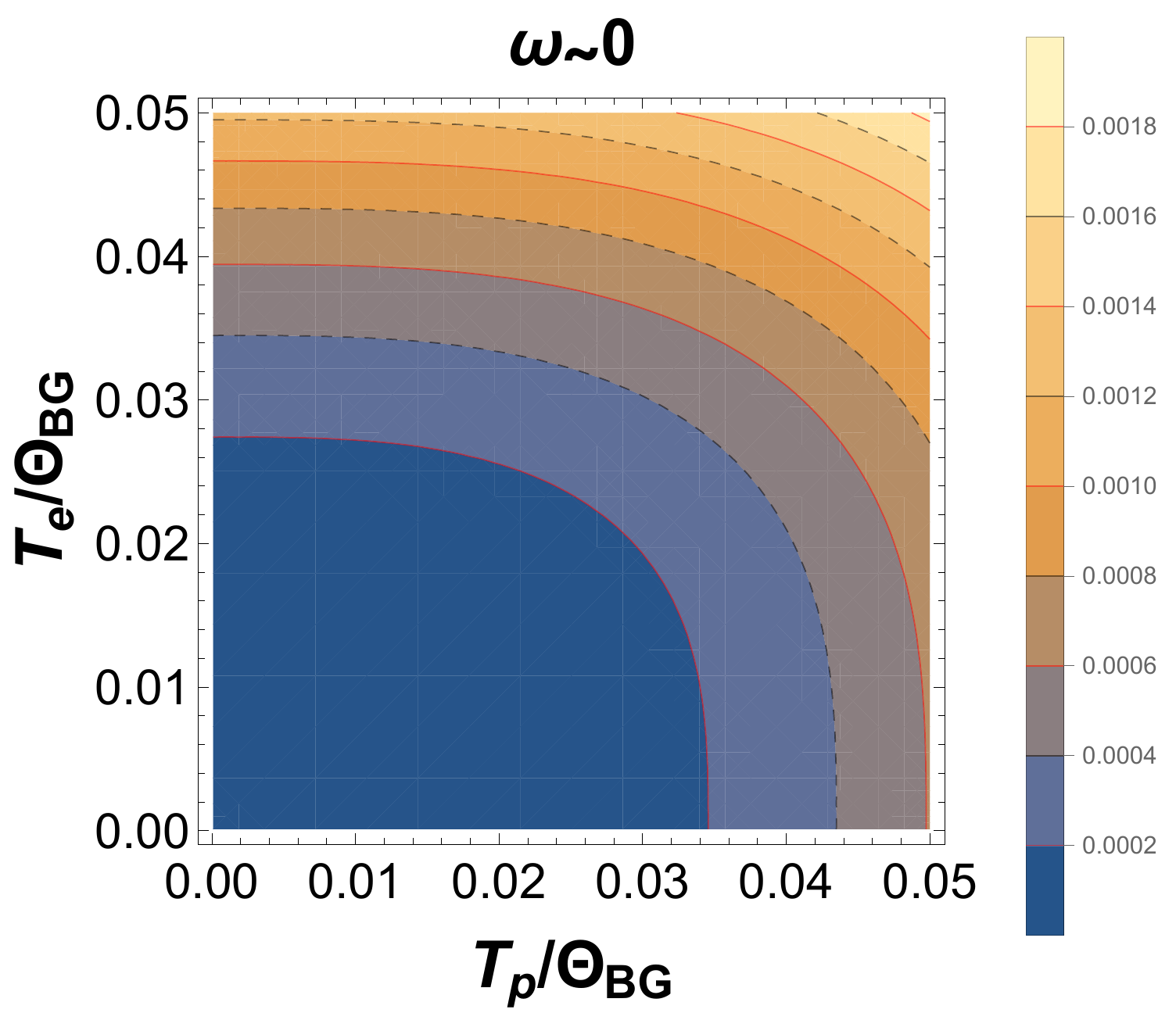}} \\
	 \subfloat[]{\includegraphics[angle=0,width=5.8cm]{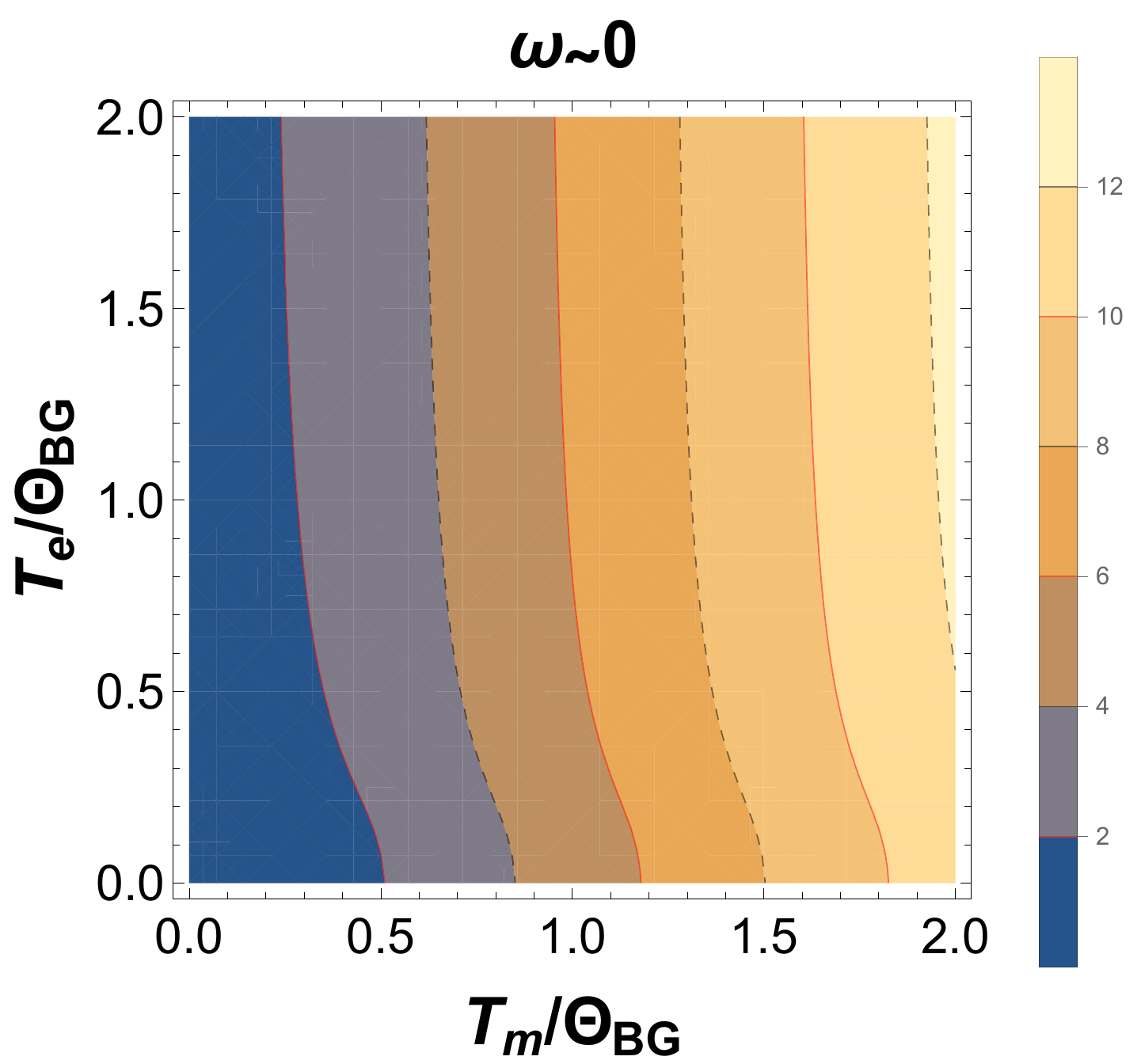}} 
	 \subfloat[]{\includegraphics[angle=0,width=5.8cm]{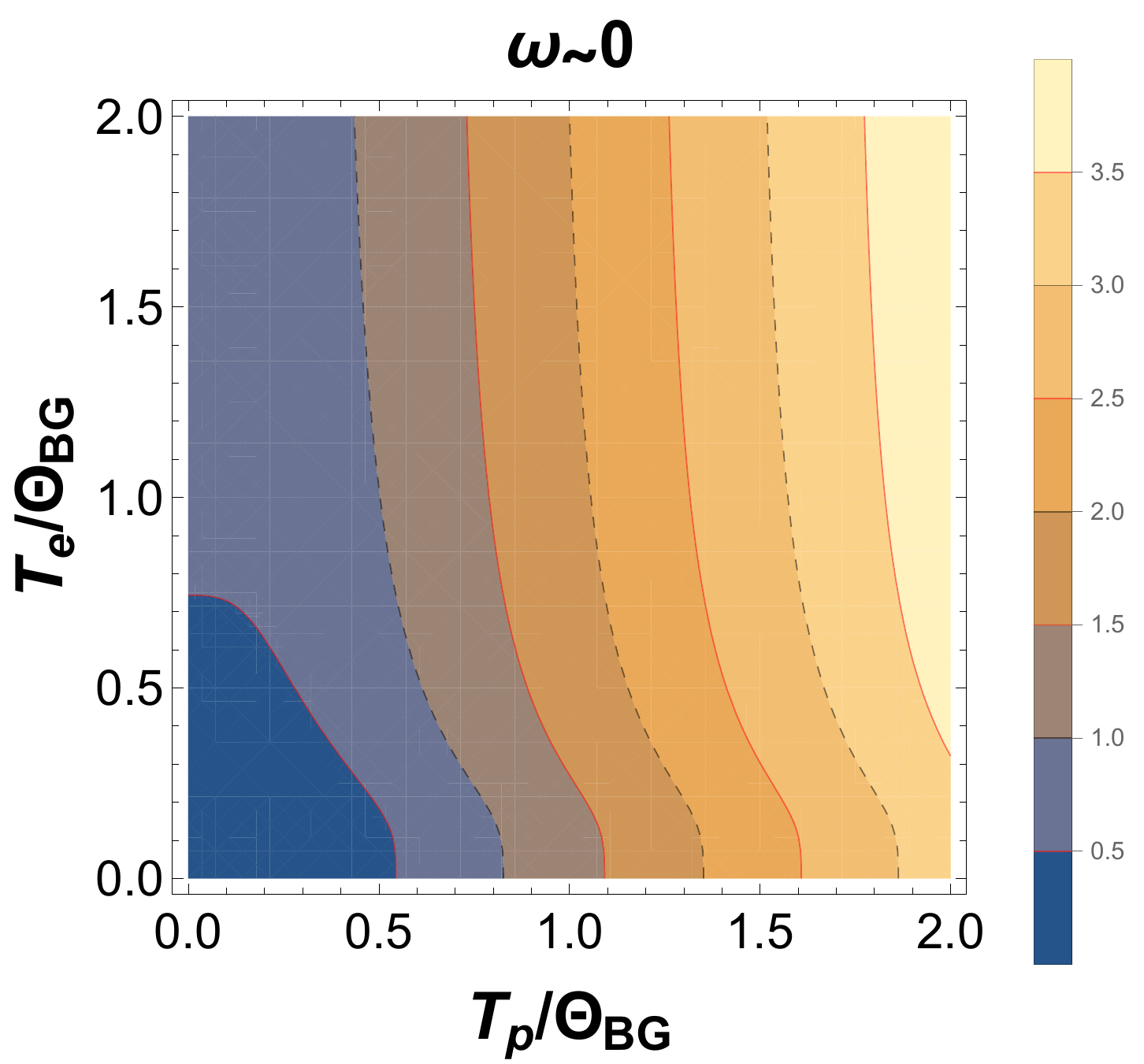}} 
\caption{(a)-(d) Variations of the temperature dependent scattering rate at zero frequency in electron-magnon interaction case and electron-phonon interaction case. (a) and (b) electron and magnon temperatures are scaled with the magnon Bloch-Gr\"{u}neisen temperature and $1/\tau (T_m,T_e)$ is normalized with $1/\tau_{e-m}^0$. Similarly, in (c) and (d), both the electron and phonon temperatures are scaled with the corresponding Bloch-Gr\"{u}neisen temperature and $1/\tau (T_p,Te)$ is scaled with $1/\tau_{e-p}^0$   for the case of electron-phonon interaction. Panels (e) and (f) show the contour plots ($T_m$ vs $T_e$ and $T_p$ vs $T_e$ ) for the scattering rate at zero frequency and lower temperature domains. (g) and (h) The high temperature regime of Figs. ((e) and (f)) are elaborated.}
	\label{fig:fig1}
\end{figure}

\begin{figure}[htbp!]
	\centering
	\subfloat[]{\includegraphics[angle=0,width=6.0cm]{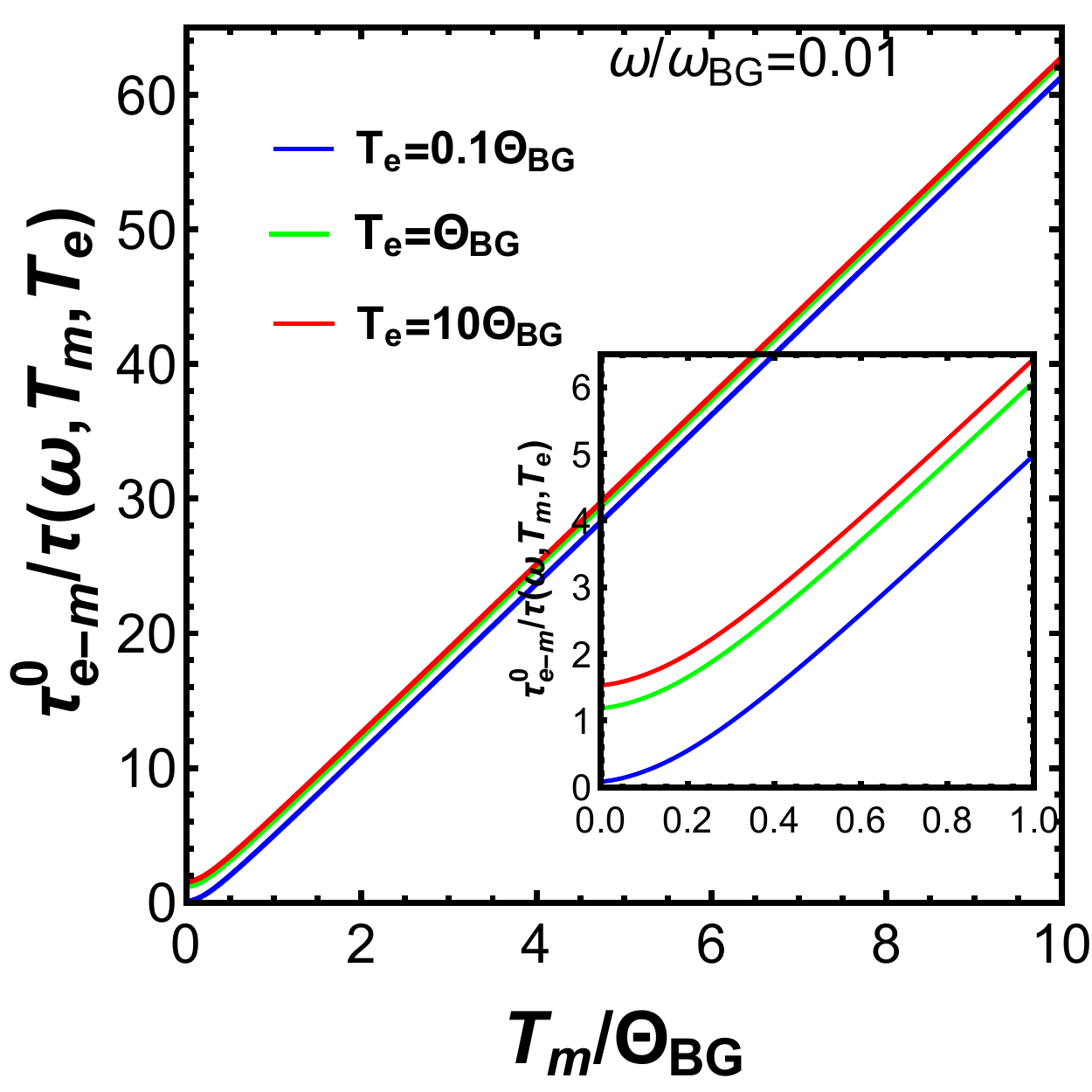}}
	\subfloat[]{\includegraphics[angle=0,width=6.0cm]{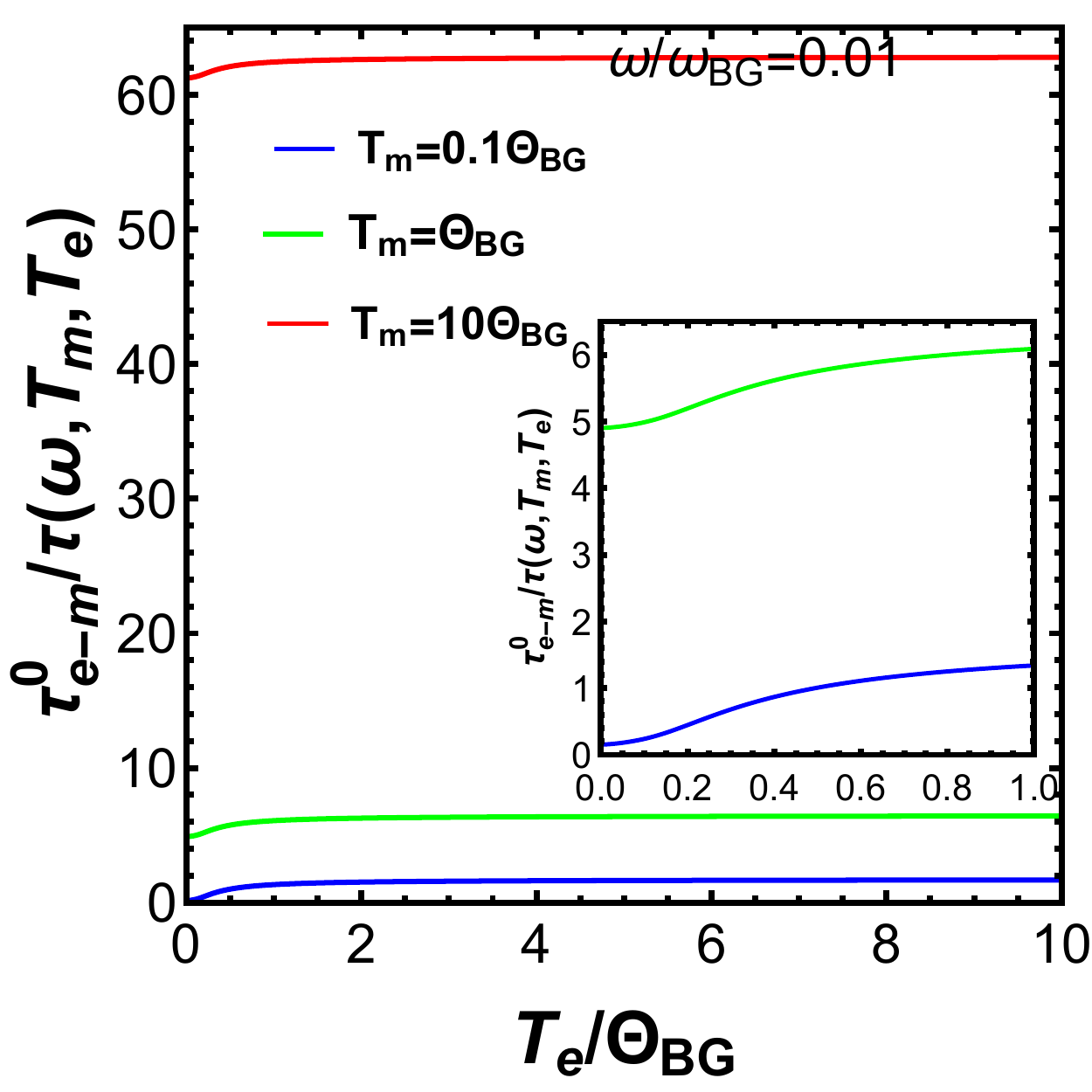}}\\
	\subfloat[] {\includegraphics[angle=0,width=6.0cm]{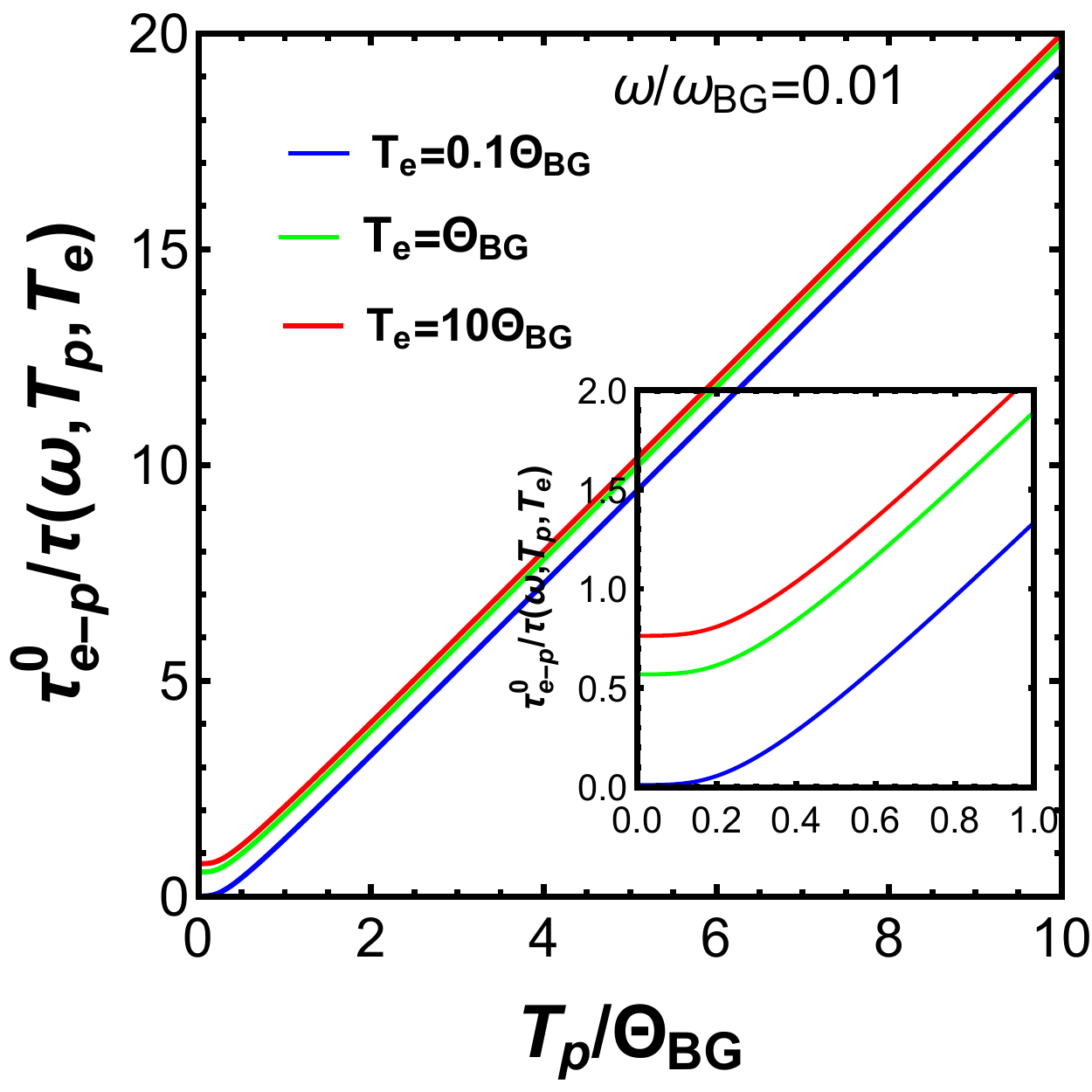}}
	\subfloat[] {\includegraphics[angle=0,width=6.0cm]{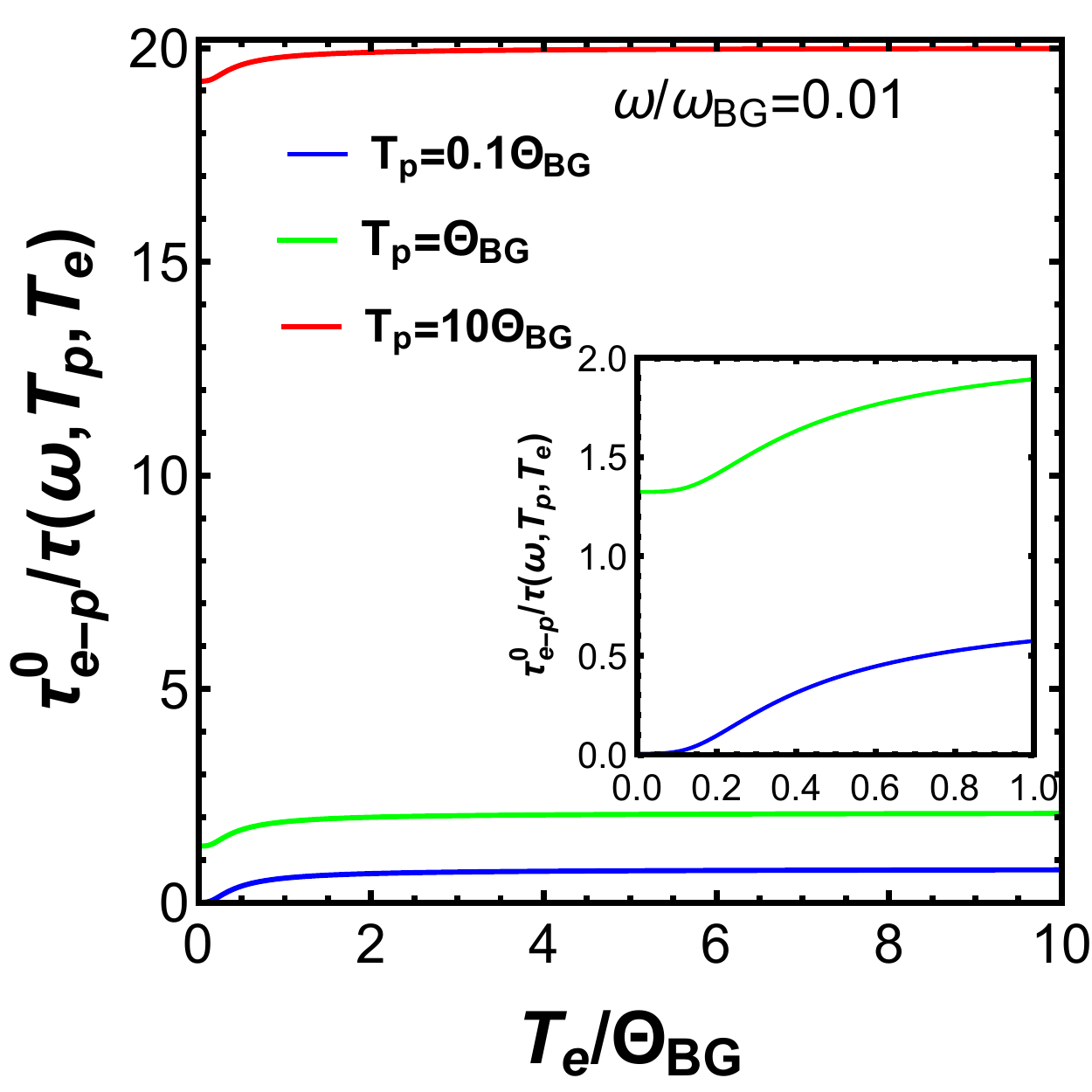}}
	\caption{Variations of the scattering rate at finite but lower frequency and different temperatures regimes for the case of electron-magnon interaction and electron-phonon interactions. The insets of Figs.(a-d), low temperature regime is elaborated, respectively. Here frequency is scaled with Bloch-Gr\"{u}neisen frequency. }
	\label{fig:fig2}
\end{figure}

\begin{figure}[htbp!]
	\centering
	\subfloat[]{\includegraphics[angle=0,width=5.7cm]{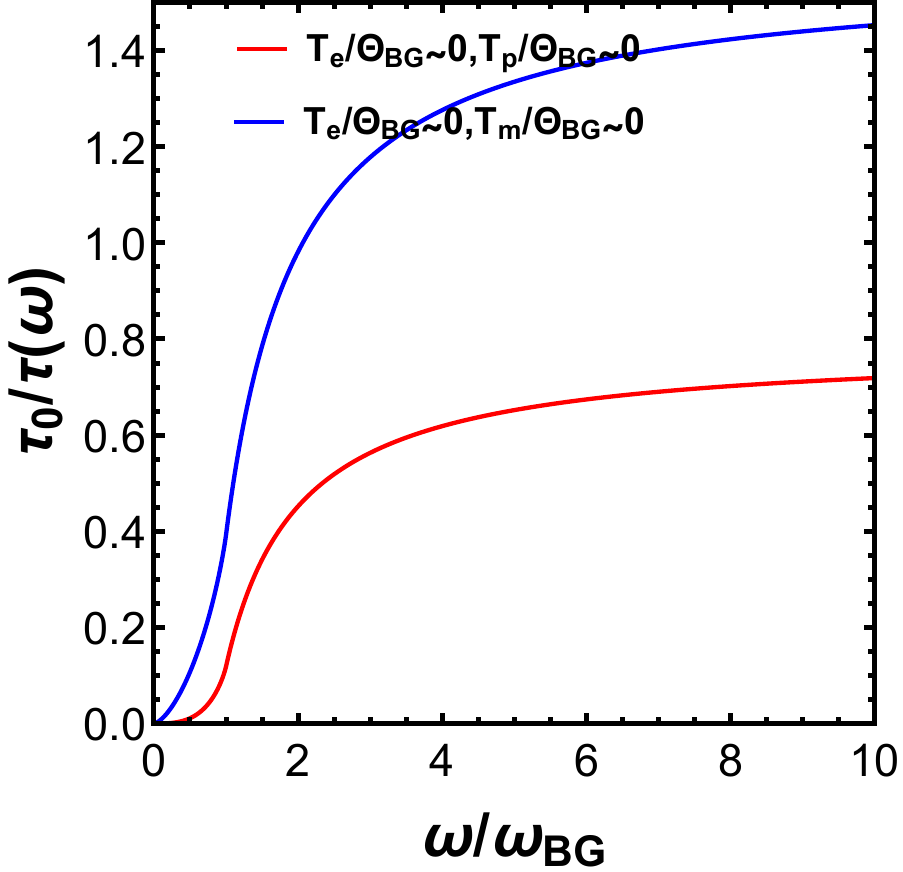}}
	\subfloat[]{\includegraphics[angle=0,width=5.7cm]{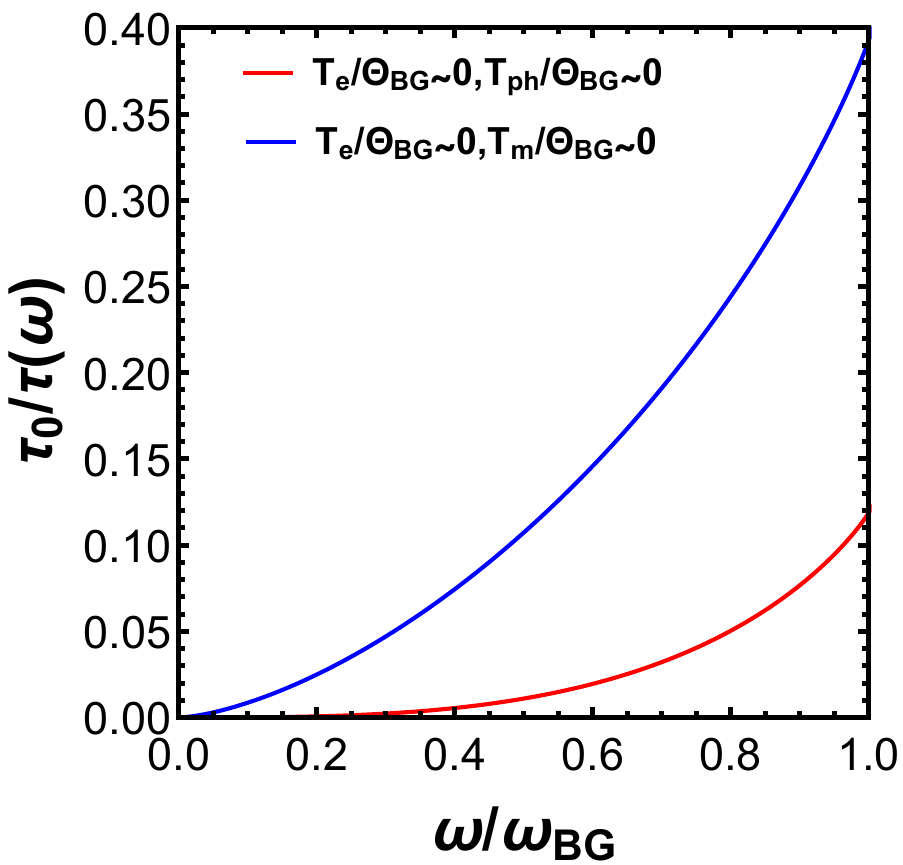}}
	\caption{(a) Variation of the scattering rate at zero temperature in electron-magnon  and electron-phonon interaction cases. (b) The low frequency domain of Fig.(a) is
		elaborated. Here $1/\tau(\w)$ is scaled with $1/\tau_0$ and frequency is scaled with Bloch-Gr\"{u}neisen frequency.}
	\label{fig:fig3}
\end{figure}

\begin{figure}[htbp!]
	\centering
	\subfloat[]{\includegraphics[angle=0,width=5.7cm]{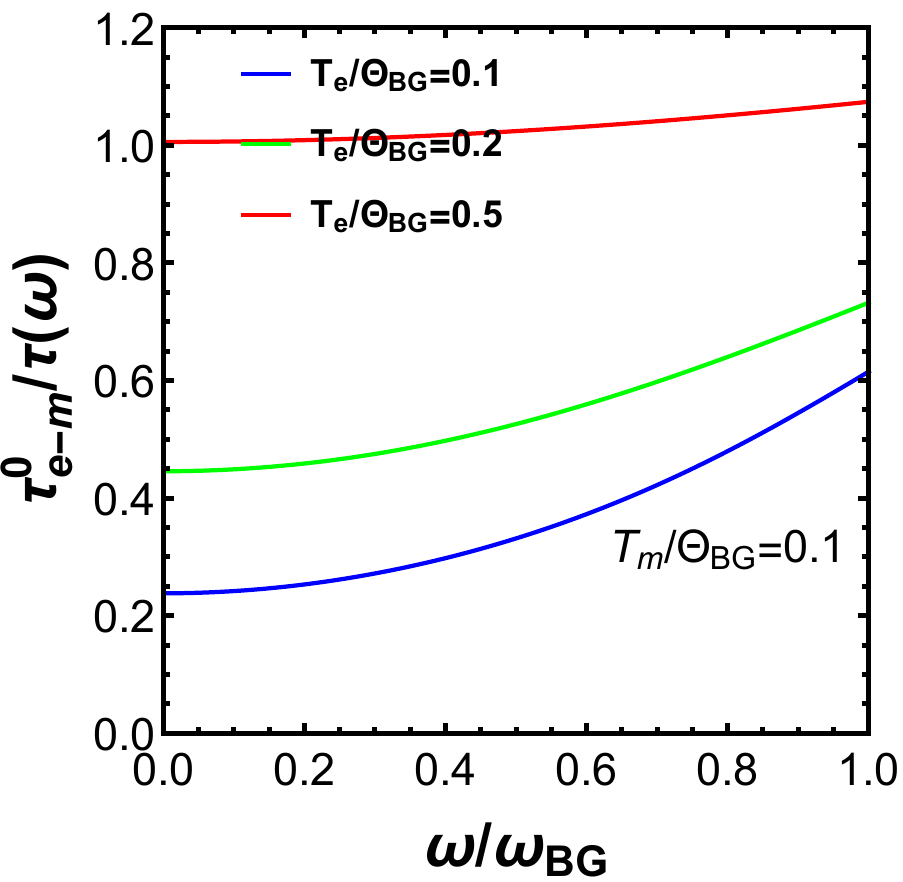}}
	\subfloat[]{\includegraphics[angle=0,width=5.7cm]{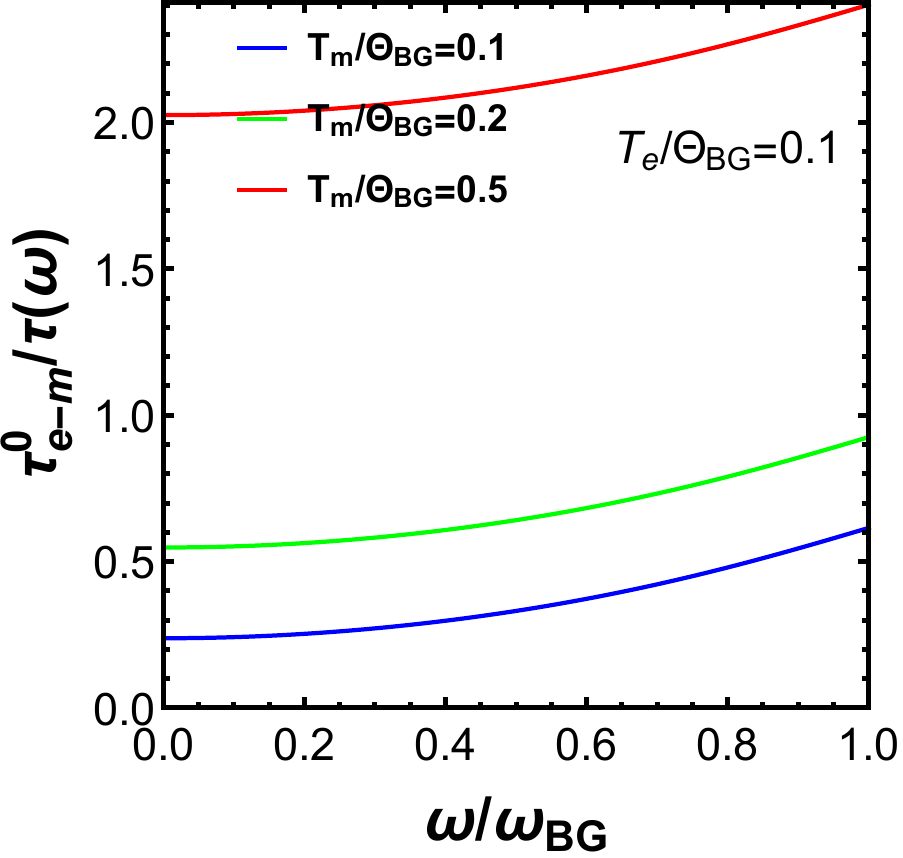}}\\
	\subfloat[]{\includegraphics[angle=0,width=5.7cm]{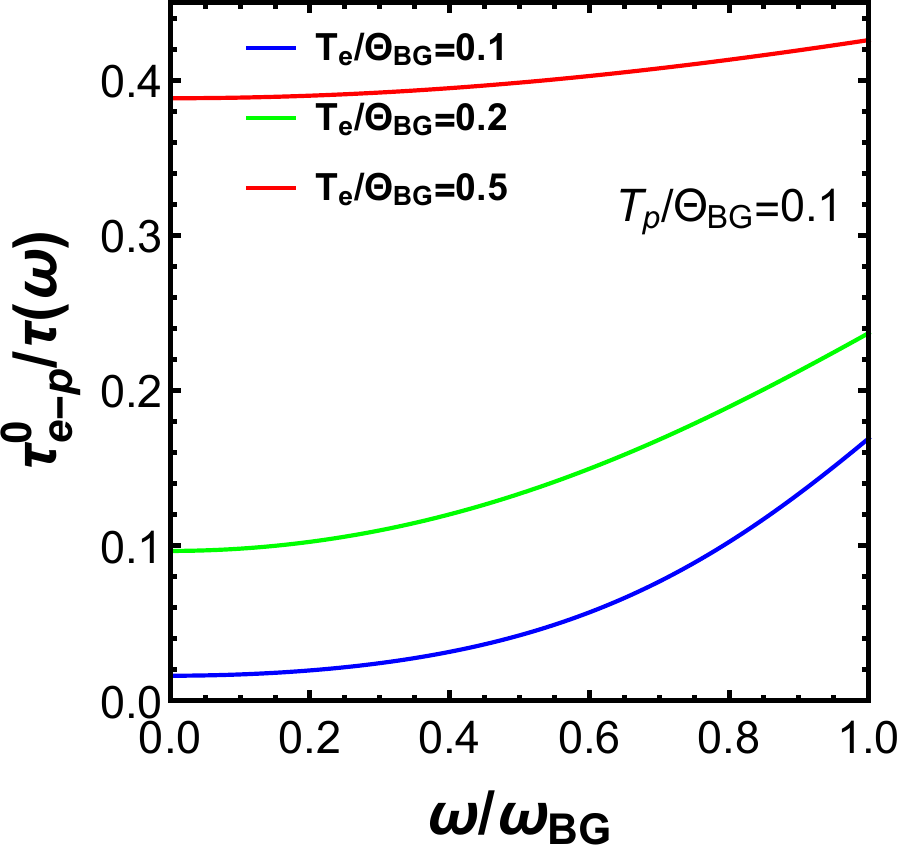}}
	\subfloat[]{\includegraphics[angle=0,width=5.7cm]{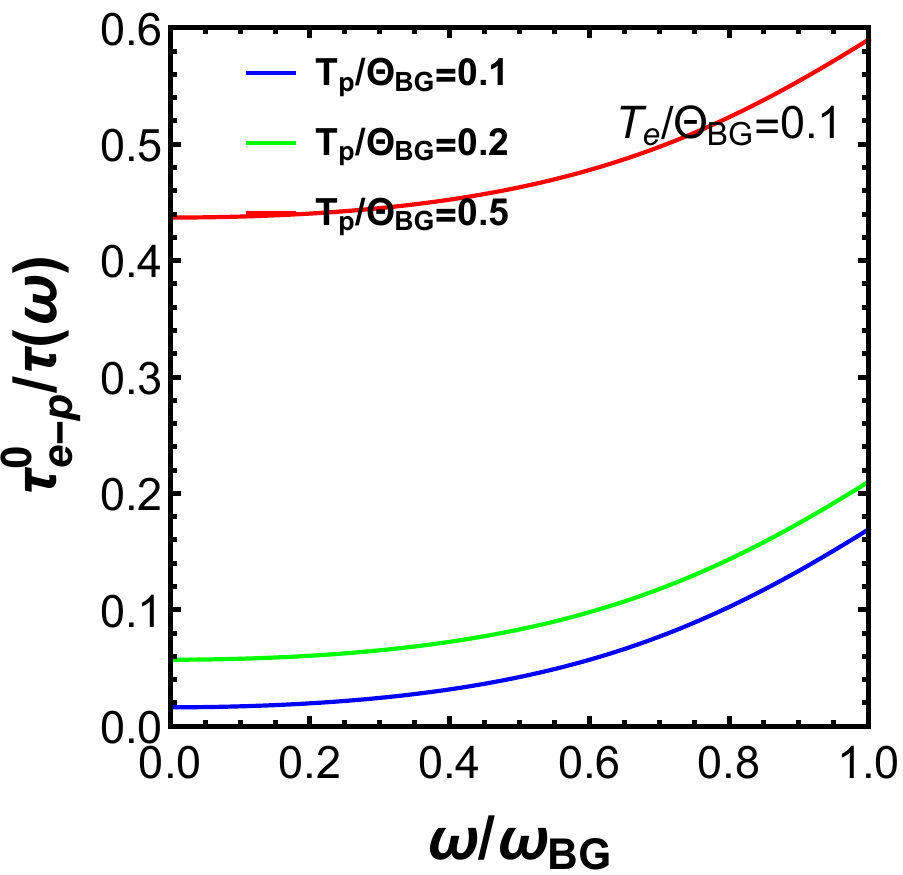}}
	\caption{ Plots of scattering rate with frequency at different temperature regimes in electron-magnon and electron-phonon interaction cases.}
	\label{fig:fig4}
\end{figure}

\end{widetext}

 \section{Summary and conclusions}
 We have presented a theoretical study of non-equilibrium relaxation of electrons due to their coupling with magnons and phonons in the normal state of iron pnictides by using the memory function approach, respectively. In zero frequency (DC) regime, we have computed scattering rate as follows:  at higher temperature ($T\gg\Theta_{BG}$) phonon and magnon scattering varies linearly with temperature while at lower temperature $T\ll\Theta_{BG}$, we have observed the different temperature scale of electron-phonon and electron-magnon scattering i.e. ($1/\tau_{e-p} \varpropto T^3$) and ($1/\tau_{e-m} \varpropto T^{3/2}$).
 
 In simple metal case, we observed that $T^5$-law of $1/\tau$ below the Debye temperature while in the normal state of iron pnictides, dc scattering rate has a $T^3$ rise below the $\Theta_{BG}$.
 In the low frequency case ($\omega\ll\omega_{D}$) and in lower temperature regimes ($T, T_e\ll\Theta_D$), the electron-phonon scattering $1/\tau_{e-p}$ in metals has three terms ($a_6T^5+b_6T_e^5+c_6T_e^5\w^2$), whereas in the corresponding case of normal state in iron pnictide this dependence changes to ($a_{10}T_p^{3} + b_{10} T_e^{3} + c_{10} \w^2 T_e$).
 
 We also predict the Holestein Mechanism\cite{TH} for the case of electron-magnon interaction and electron-phonon interaction in the normal metallic state of iron-based pnictides. For the electron-magnon study, in zero temperature limit ($T_m, T_e \rightarrow 0$) and at lower frequency, we observe that the scattering rate scales as $\w^{3/2}$ and at frequency higher than $\Theta_{BG}$ it saturates to a constant value. For the case of electron-phonon interaction, in zero temperature behavior and lower frequency regime ($\omega\ll\omega_{BG}$), it is notice that the scattering rate proportional to $\w^3$ power law in contrast to the simple metal (scales as $\w^5$)\cite{GW} and graphene (scales as $\w^4$)\cite{LR}.   
 
 Also, notice that the DC scattering rate at high temperatures and the AC scattering rate at higher frequencies over BG frequency are independent of the temperature difference
 between the electrons and the bosons (magnons or phonons).  These results can be viewed in a pump-probe spectroscopic experiments.

 Further, it will be interesting to extend these calculations by taking multi-orbital coupling along with magonon and phonon baths for better understanding the transport phenomenon in the normal state and superconducting state of iron-based pnictides in the steady state nonequilibrium situation. 
 \\

  {\textbf{Acknowledgment}} \\
  
  We thank to Navinder Singh and Haranath Ghosh for many useful discussions.
  One of the authors (L.R) is supported by the Scientific and Technical Research Council of Turkey (T$\ddot{U}$BITAK) ARDEB International project no 118F187. 


\end{document}